\documentclass[%
 aip,
 amsmath,amssymb,
 reprint,%
]{revtex4-1}

\usepackage{graphicx}
\usepackage{dcolumn}
\usepackage{bm}

\usepackage[utf8]{inputenc}
\usepackage[T1]{fontenc}
\usepackage{mathptmx}
\usepackage{etoolbox}

\usepackage{siunitx}
\usepackage{subcaption} 
\usepackage{xcolor}
\usepackage{mathtools}
\usepackage{amsmath}
\usepackage{amssymb}
\usepackage[justification=justified]{caption}
\usepackage{float}
\usepackage{booktabs, makecell, tabularx}
\usepackage[strict]{changepage}

\makeatletter
\renewenvironment{subarray}[2][c]{%
  \if#1c\vcenter\else\vbox\fi\bgroup
  \Let@ \restore@math@cr \default@tag
  \baselineskip\fontdimen10 \scriptfont\tw@
  \advance\baselineskip\fontdimen12 \scriptfont\tw@
  \lineskip\thr@@\fontdimen8 \scriptfont\thr@@
  \lineskiplimit\lineskip
  \ialign\bgroup\ifx c#2\hfil\fi
    $\m@th\scriptstyle##$\hfil\crcr
}{%
  \crcr\egroup\egroup
}
\makeatother
\renewcommand{\substack}[2][c]{\subarray[#1]{c}#2\endsubarray}

\newcommand {\arraystretchdefaultl} {1.3}
\newcommand{\arraystretchdefault}{ \renewcommand {\arraystretch} {\arraystretchdefaultl} }
\arraystretchdefault

\DeclareFontFamily{U}{mathc}{}
\DeclareFontShape{U}{mathc}{m}{it}%
{<->s*[1.03] mathc10}{}

\DeclareMathAlphabet{\mathscr}{U}{mathc}{m}{it}

\newcommand{\be}[1]{\begin{equation}\label{#1}}
\newcommand{\ee}{\end{equation}}
\newcommand{\ba}[1]{\begin{eqnarray}\label{#1}}
\newcommand{\ea}{\end{eqnarray}}
\newcommand{\rf}[1]{(\ref{#1})}
\newcommand{\nn}{\nonumber}
\newcommand{\dd}{{\rm{d}}}

\makeatletter
\def\@email#1#2{%
 \endgroup
 \patchcmd{\titleblock@produce}
  {\frontmatter@RRAPformat}
  {\frontmatter@RRAPformat{\produce@RRAP{*#1\href{mailto:#2}{#2}}}\frontmatter@RRAPformat}
  {}{}
}%
\makeatother
\begin{document}

\preprint{AIP/123-QED}

\title[{\color{black}{Diffusive instabilities of baroclinic lenticular vortices}}]{{\color{black}{Diffusive instabilities of baroclinic lenticular vortices}}}
\author{Joris Labarbe}
 \email{joris.labarbe@northumbria.ac.uk}
\affiliation{Institut de Recherche sur les Ph\'enom\`enes Hors Equilibre, UMR 7342, CNRS - Aix-Marseille
Universit\'e, 49 rue F. Joliot Curie, 13384 Marseille, CEDEX 13, France}
\affiliation{ 
Northumbria University, Newcastle upon Tyne NE1 8ST, UK
}
\author{Oleg N. Kirillov}%
 \email{oleg.kirillov@northumbria.ac.uk}
\affiliation{ 
Northumbria University, Newcastle upon Tyne NE1 8ST, UK
}%

\date{\today}

\begin{abstract}
We consider a model of a circular lenticular vortex immersed into a deep and vertically stratified viscous fluid in the presence of gravity and rotation. The vortex is assumed to be baroclinic with a Gaussian profile of angular velocity both in the radial and axial directions. Assuming the base state to be in a cyclogeostrophic balance, we derive linearized equations of motion and seek for their solution in a geometric optics approximation to find amplitude transport equations that yield a comprehensive dispersion relation. Applying algebraic Bilharz criterion to the latter, we establish that stability conditions are reduced to three inequalities that define stability domain in the space of parameters. The main destabilization mechanism is either {\color{black}{monotonic}} or oscillatory axisymmetric instability depending on the Schmidt number ($Sc$), vortex Rossby number and the difference between the radial and axial density gradients as well as the difference between the epicyclic and vertical oscillation frequencies. We discover that the boundaries of the regions of {\color{black}{monotonic}} and oscillatory axisymmetric instabilities meet at a codimension-2 point, forming a singularity of the neutral stability curve. We give an exhaustive classification of the geometry of the stability boundary, depending on the values of the Schmidt number. Although we demonstrate that the centrifugally stable (unstable) Gaussian lens can be destabilized (stabilized) by the differential diffusion of mass and momentum and that destabilization can happen even in the limit of vanishing diffusion, we also describe explicitly a set of parameters in which the Gaussian lens is stable for all $Sc>0$.
\end{abstract}

\maketitle

%

\section{Introduction}

\begin{figure}
\centering
\includegraphics*[width=.45\textwidth]{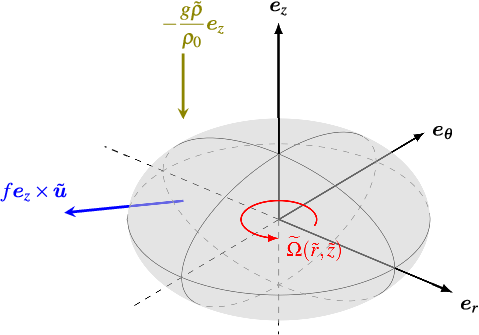}

\caption{Sketch of a differentially rotating lenticular vortex in a cyclogeostrophic balance between centrifugal, hydrostatic and Coriolis forces. \label{sketch}}
\end{figure}

An intriguing class of dynamical systems of geophysics and astrophysics resides in the so-called \textit{lenticular vortices} that serve to model mesoscale oceanic or atmospheric cyclons and anticyclons, such as the Great Red Spot (GRS) of Jupiter \cite{EGCL2020,BM2005,LU2019}. These compact but intense three-dimensional baroclinic vortices are strongly influenced by planetary rotation and thus are governed by geostrophic and hydrostatic balances between pressure gradients, Coriolis, and buoyancy forces, from where they get their ellipsoidal shape \cite{YSB2019}, see Fig.~\ref{sketch}. The aspect ratio of the vertical half-height to the horizontal length scale of such vortices in an equilibrium state ranges from flat ‘pancakes’ to nearly round and depends on the properties of both the ambient flow and the vortex \cite{Au2012,HML2012}. 

The observed in geophysics lenticular vortices are notoriously persistent (like the GRS). In particular, a relatively long life cycle of weeks to years allows the intense oceanic eddies to transport heat, salt, and other passive tracers over long distances and thus  to contribute to the climate equilibrium on Earth \cite{Z2017}. Nevertheless, even the GRS is subject to variations in its size \cite{GRS2018}. Therefore, natural and timely questions arise on how stable such lenticular vortices are, what are their basic destabilization mechanisms, how quickly are they decaying, and what are their origins? 

Indeed, although according to the Taylor-Proudman theorem rotation tends to generate tall barotropic columnar vortices, many studies show that vortices in rotating and stably stratified fluids  have a lenticular shape  rather than being columnar \cite{YB2016}. There is an evidence that lenticular vortices in rotating and stratified fluids can
be created, e.g., from the remnants of the violent breakups of columnar vortices \cite{HML2012} or due to zigzag instability that destroys interacting columnar vortices \cite{YB2016}. 

Among the main instabilities of single columnar axisymmetric vortices such as shear instability, centrifugal instability, radiative instability, and Gent–McWilliams instability\cite{GM1986}, the latter was found to be the most effective in bending and slicing the vortex into lenticular vortices embedded within shallow layers as it happens in many geophysical flows \cite{YB2015,Au2012,Z2017,L2013}. 

We notice that coherent vortices generated from wakes or turbulence in stratified fluids or through hydrodynamic instabilities of surface currents can also have an ellipsoidal shape \cite{YB2016}.

{\color{black}{Monotonic}} axisymmetric modes subjected to centrifugal instability are frequently found to be the most unstable in numerical, theoretical and experimental works on stability of vortices \cite{YB2016,YSB2019}. The onset of the centrifugal instability for inviscid differentially rotating axisymmetric circular non-stratified flows, including columnar (barotropic) vortices, is regulated by the standard Rayleigh criterion (1917), requiring the square of the absolute angular momentum to decrease with the radius somewhere in the flow \cite{YB2016}. 

The Rayleigh criterion was extended in numerous works taking into account, accordingly, rotation effects \cite{M1992}, non-axisymmetric disturbances \cite{BG2005}, vertical stratification \cite{S1936,SH2013}, temperature gradient \cite{SH2013,KM2017,BJT2020}, magnetic field and other physical phenomena \cite{AG78,KSF2014,O2016}. However, it was widely accepted that even the generalized inviscid Rayleigh criterion cannot be adequately applied to oceanic eddies unless diffusion of momentum (viscosity) and of a stratifying agent (diffusivity) is added to the model.

For instance, Lazar et al. \cite{L2013} considered different types of circular barotropic
vortices in a linearly stratified shallow layer of viscous fluid with the ratio between the kinematic viscosity and mass diffusivity (i.e. Schmidt number, $Sc$) fixed to unity and found that the centrifugally unstable area in the parameter space is reduced in the double-diffusive setting with respect to the diffusionless one. Besides, Lazar et al. \cite{L2013} provided an analytical marginal
stability limit for the idealized Rankine vortex in terms of the vortex Rossby number and Burger and Ekman numbers.

The effect of the Schmidt number on the stability of barotropic vortices in a stratified ambient fluid in the absence of rotation have been studied recently by Singh and Mathur \cite{SM2019} with the geometric optics approach developed in \cite{KSF2014,KM2017,K2017,K2021}.

Lenticular vortices of geophysical interest are generally baroclinic with their 
azimuthal velocity being a function of both radial and axial coordinate. Nevertheless, many previous studies addressing stability of baroclinic vortices, concentrated on the extension of the Rayleigh centrifugal criterion for barotropic vortices. For instance, the generalized Rayleigh criterion for baroclinic and stratified
circular flows proposed by Solberg (1936) states that the flow is unstable if the total circulation decreases as the radius increases along isopycnals in the flow \cite{S1936,SH2013,YSB2019,YB2016}.

In his seminal work, McIntyre addressed stability of
a baroclinic circular vortex {\color{black}{in a Boussinesq fluid with buoyancy determined by the temperature gradient only}} and demonstrated that even in the limit of vanishing diffusivities of momentum and heat, the centrifugally stable diffusionless vortex is actually unstable to axisymmetric disturbances, unless the Prandtl number, which is the ratio of viscosity to thermal diffusivity, is equal to unity \cite{M1970}. Furthermore, `the more the Prandtl number differs from 1, the larger the region in parameter space for which the flow is stable by the classical criterion, but actually unstable' \cite{M1970}. Although both monotonic instability and growing oscillations have been found, only the former was shown to determine the stability criterion \cite{M1970}. 

Despite the lenticular vortices are very common in stratified fluids, the literature on their stability is not vast  \cite{B2001,GC2003,YB2016,YSB2019,Au2012,HML2012,Z2017,EGCL2020,EBM2016,GCB2004,LB2021}. The vast majority of these works are purely numerical and experimental. Therefore, a general analytical treatment of this problem is timely to get new tools for interpreting the data, informing, and guiding further research.

Laboratory experiments demonstrated  that most of the laboratory lenticular vortices are unstable to baroclinic or barotropic instabilities depending on their aspect
ratio \cite{Z2017}.

Beckers et al. \cite{B2001} and Godoy-Diana and Chomaz \cite{GC2003} have studied the effect of the Schmidt number on the decay of axisymmetric pancake vortices in a stratified fluid. It was found that for $Sc>1$ the secondary circulation inside the vortex,
dominated by the diffusion of momentum, slows down the decay of the horizontal velocity whereas for $Sc<1$ the secondary motion, primarily driven
by the density diffusion, accelerates the damping of the velocity \cite{GC2003}.

Yim et al. \cite{YSB2019} proposed a model incorporating a multitude of different velocity profiles in radial and axial directions for the vortices, including Gaussian-Gaussian, Gaussian-columnar and Gaussian-exponential ones. However, the Schmidt number was fixed to
unity in that work, `since the turbulent advection at small scales dominates the molecular viscosity and diffusivity'.

Yim and Billant \cite{YB2016} explored the difference and similarity in destabilization of columnar and ellipsoidal vortices in a non-rotating but double-diffusive setting. Despite in most of the paper \cite{YB2016}, $Sc = 1$ was kept for simplicity as well, the effect of the Schmidt number was briefly investigated in it. The authors found numerically a new instability branch after increasing $Sc$ to 700. This branch corresponded to inclined short-wavelength oscillations localized in the top and bottom of the vortex and was attributed by Yim and Billant \cite{YB2016} to the oscillatory McIntyre instability \cite{M1970}. Nevertheless, since this instability was found to co-exist with the centrifugal instability, being `less unstable', it was not investigated further by Yim and Billant \cite{YB2016}. {\color{black}{Recent numerical study \cite{LB2021}, however, provides new evidence that the McIntyre instability is a reason for density layer formation observed around laboratory \cite{GL1981} and oceanic \cite{HMEA2013} vortices.}}

In the present work our ambition is: (i) to derive an original set of dimensionless equations for circular baroclinic lenticular vortices in a vertically stratified and rotating ambient fluid taking into account diffusion of momentum and mass; (ii) to perform a local stability analysis within the geometric optics approach \cite{KSF2014,KM2017,K2017,K2021}; (iii) to find a comprehensive dispersion relation allowing stability analysis with arbitrary values of $Sc$; (iv) to find analytically new explicit instability criteria generalizing the previous results.

We will show that the Gaussian-Gaussian lenticular vortex is subject to both the {\color{black}{monotonic}} axisymmetric centrifugal instability and oscillatory McIntyre instability depending on the value of the Schmidt number. We will show that it is a codimension-2 point on the neutral stability curve that separates the criteria for the centrifugal instability and for the oscillatory McIntyre instability. Finally, we will provide evidence that the latter is a genuine double-diffusive dissipation-induced instability, which criterion exists only when $Sc\ne 1$, and which persists even in the limit of vanishing diffusion.


\section{Mathematical setting}

Let $\tilde{t}$ be time and $(\tilde{r},\tilde{\theta},\tilde{z})$ be a right-handed cylindrical coordinate system with the unit vectors $\bm{e}_r$, $\bm{e}_{\theta}$, $\bm{e}_z$, Fig.~\ref{sketch}. We assume that the frame of reference rotates with angular velocity $(0, 0, {\color{black}{f/2}})$, where the constant Coriolis {\color{black}{parameter}}, $f$, can be of both positive and negative sign. Gravity $(0, 0, -g)$ is anti-parallel to $\tilde z$-axis and the centrifugal force
is assumed to be negligible \cite{M1970}.

We consider a base state of a linearly stratified fluid along the direction of application of gravity. We also include dissipation in the fluid in the form of viscosity and we assume {\color{black}{for definiteness}} the diffusion of stratifying agent to be present, {\color{black}{in contrast to the work of McIntyre\cite{M1970}, where thermal diffusion only was taken into account.}} 

We further assume a baroclinic ellipsoidal vortex with the angular velocity $(0,0,\tilde \Omega(\widetilde r, \tilde z))$ to be immersed in a deep and  motionless (in the rotating frame) fluid far away from the core center at $(\tilde{r}_0,\tilde{z}_0)=(0,0)$, so that boundaries do not influence the inner motion, Fig.~\ref{sketch}.

{\color{black}{We notice, however, that the thermal diffusivity excceds by two orders of magnitude the mass diffusivity in oceans. Therefore, introducing  thermal diffusivity and the associated Prandtl number to model instabilities of a truly thermohaline vortex is a natural though challenging extension of our study, which is left for a future work. }}

\subsection{Density stratification}

The linear density variation in the vertical direction $\tilde z$ is described within the \textit{Boussinesq approximation} by the stable background density gradient $- {\rho_0 N^2}/{g}$, where $N = \sqrt{-(g/\rho_0)(\dd \tilde{\rho}/\dd \tilde{z})}$ is the Brunt-V{\"a}is{\"a}l{\"a} frequency of the ambient fluid and  $\rho_0$ is the constant reference density. 

The influence of the internal stratification of the baroclinic vortex is
captured by the density \textit{anomaly} term $\tilde{\rho}_A$ such that the total
density $\tilde{\rho}$ takes the form \cite{DV2003,YSB2019,EBM2016,BGC2021}
\be{rho}
\tilde{\rho} (\tilde{r},\tilde{z})  = \rho_0 - \rho_0\frac{ N^2}{g}\tilde{z} + \tilde{\rho}_A(\tilde{r},\tilde{z}).
\ee

\subsection{Dimensional equations of motion on the $f$-plane}

Equations on the $f$ - plane that govern evolution of the velocity field $\tilde{\bm{u}}$, density $\tilde{\rho}$, and pressure $\tilde{P}$ represent conservation of linear momentum (the Navier-Stokes equations), conservation of density, and incompressibility of the fluid:

\begin{subequations}
\label{eom}
\begin{align}
\frac{\partial \tilde{\bm{u}}}{\partial \tilde{t}} + \left( \tilde{\bm{u}} \cdot \bm{\nabla} \right) \tilde{\bm{u}} + f \bm{e}_z \times \tilde{\bm{u}} &= - \frac{\bm{\nabla} \tilde{P}}{\rho_0} - \frac{g\tilde{\rho}}{\rho_0} \bm{e}_z + \nu \bm{\nabla}^2 \tilde{\bm{u}} , \label{eom1} \\
\frac{\partial\tilde{\rho}}{\partial \tilde{t}} + \left( \tilde{\bm{u}} \cdot \bm{\nabla} \right) \tilde{\rho} &= \kappa \bm{\nabla}^2 \tilde{\rho} , \label{eom2} \\
\bm{\nabla} \cdot \tilde{\bm{u}} &= 0 . \label{eom3}
\end{align}
\end{subequations}
Here $\bm{e}_z$ is the unit vector of the chosen coordinate frame, $g$ stands for the uniform gravity acceleration and $\nu$ and $\kappa$ are the coefficients of kinematic viscosity and diffusivity, respectively \cite{YSB2019}.

It is instructive to re-write system \rf{eom} by projecting the equations onto the vertical direction, $\bm{e}_z$, and the horizontal direction specified with the vector $\bm{e}_h$ that {\color{black}{lies}} in the plane spanned by the vectors $\bm{e}_r$ and $\bm{e}_{\theta}${\color{black}{, as implemented in previous articles\cite{GC2003,GCB2004}. This transformation retains the cylindrical geometry with the curvature terms still present on the horizontal plane but it simplifies further the system and allows us to perform a thorough dimensional analysis of the variables. Equations \rf{eom} thus become}}
\begin{subequations}
\label{eom2}
\begin{align}
\frac{\partial \tilde{\bm{u}}_h}{\partial \tilde{t}} + \left( \tilde{\bm{u}}_h \cdot \widetilde{\bm{\nabla}}_h \right) \tilde{\bm{u}}_h + \tilde{u}_z \frac{\partial \tilde{\bm{u}}_h}{\partial \tilde{z}} + f \bm{e}_z \times \tilde{\bm{u}}_h &= - \frac{1}{\rho_0} \widetilde{\bm{\nabla}}_h \tilde{P}  \nn \\ &+\nu \widetilde{\mathcal{D}} \tilde{\bm{u}}_h , \label{eom2-1} \\
\frac{\partial \tilde{u}_z}{\partial \tilde{t}} + \left( \tilde{\bm{u}}_h \cdot \widetilde{\bm{\nabla}}_h \right) \tilde{u}_z + \tilde{u}_z \frac{\partial \tilde{u}_z}{\partial \tilde{z}} &= - \frac{1}{\rho_0} \frac{\partial \tilde{P}}{\partial \tilde{z}}  \nn \\ &-\frac{g\tilde{\rho}}{\rho_0} + \nu \widetilde{\mathcal{D}} \tilde{u}_z , \label{eom2-2} \\
\frac{\partial\tilde{\rho}}{\partial \tilde{t}} + \left( \tilde{\bm{u}}_h \cdot \widetilde{\bm{\nabla}}_h \right) \tilde{\rho} + \tilde{u}_z \frac{\partial \tilde{\rho}}{\partial \tilde{z}} &= \kappa \widetilde{\mathcal{D}} \tilde{\rho} , \label{eom2-3} \\
\widetilde{\bm{\nabla}}_h \cdot \tilde{\bm{u}}_h +  \frac{\partial \tilde{u}_z}{\partial \tilde{z}} &= 0 , \label{eom2-4}
\end{align}
\end{subequations}
where $\widetilde{\mathcal{D}}$ is the operator defined as
\be{opD}
\widetilde{\mathcal{D}} = \widetilde{\bm{\nabla}}_h^2 + \frac{\partial^2}{\partial \tilde{z}^2}.
\ee

\subsection{Non-dimensionalization}

Let us introduce scaling laws as follows
\ba{nondim}
&\tilde{r} = r^{*}r, \quad  \tilde{\theta}=\theta, \quad \tilde{z} = z^{*}z, \quad \tilde{t} = t^{*} t, \quad \tilde{\bm{u}} = u_h^{*} \bm{u}_h  + u_z^{*} u_z \bm{e}_z,& \nn \\
&\tilde{\rho} = \rho^{*} \rho , \quad \tilde{P} = P^{*} P,&
\ea
where $t^{*} = R/U$ is an advective time scale, $(r^{*},z^{*}) = (R,Z)$ are characteristic radial and axial length scales and $(u_h^{*},u_z^{*}) = (U,W)$ are typical horizontal and vertical velocities, which we assume to be positive. {\color{black}{We emphasize that viscous diffusion is neglected in the base flow of the vortex in a manner that the radius $R$ does not evolve according to the scaling law\cite{RDM2010} $R=\sqrt{R_0^2 + 4\nu t}$ but instead, remains constant over time (as it is the case if one assumes $\nu\to 0$).}}

Dividing equation \rf{eom2-1} by the factor $fU$, we obtain the scaling law for pressure as being
\be{ndp}
P^{*} = \rho_0 f R U.
\ee
We use a similar methodology to recover the dimensional factor $\rho^{*}$ for the density, from the balance between {\color{black}{non-hydrostatic}} pressure and buoyancy forces in expression \rf{eom2-2}, yielding
\be{ndp}
\rho^{*} = \frac{P^{*}}{gZ} = \frac{\rho_0 f U}{g \alpha} ,
\ee
while introducing the aspect ratio of the vortex
\be{alpha}
\alpha = \frac{Z}{R} .
\ee

Finally, we make use of expression \rf{eom2-3} to recover the scaling law for the vertical velocity $W$. Substituting the previous factors and density profile \rf{rho} in this equation without presence of diffusivity ($\kappa=0$) yields the following balance
\be{cod}
\left( \frac{W \rho_0 R N^2}{g U \rho^{*}} \right) u_z = \left( \frac{W}{U \alpha} \right) u_z \frac{\partial \rho_A}{\partial z}.
\ee
From expression \rf{cod}, two distinct scaling laws are possible for the axial velocity $W$, namely $W \sim (gU\rho^{*})/(\rho_0RN^2)$ or $W \sim U\alpha$, depending on the regime considered (strong or weak stratification and rotation rate). We further introduce, respectively, the horizontal Froude number as the ratio of the flow velocity over the maximum phase speed of internal gravity waves \cite{GC2003,GCB2004,BT2013} and the vortex Rossby number as the ratio of the angular velocity of the vortex to the Coriolis frequency \cite{YSB2019}
\be{FhRo}
F_h = \frac{U}{RN}, \quad Ro = \frac{U}{fR}.
\ee
As $U>0$, anticyclonic (cyclonic) eddies correspond to negative (positive) values of $Ro$ and $f$ \cite{YSB2019}.

In the following we assume a reasonable for the geophysical applications regime with strong stratification and large in absolute value Coriolis parameter $f$, such that the ratio $F_h^2/Ro$ is of order unity and thus, both scales for $W$ are consistent whatever the value of $\alpha$ is \cite{EBM2016}. We therefore choose $W=\alpha U$ for the sake of simplicity of the equations of motion.

To complete the set of dimensionless parameters of consideration, we introduce two more dimensionless numbers, namely the Schmidt and the Ekman numbers
\be{ScEk}
Sc = \frac{\nu}{\kappa} \quad \textrm{and} \quad Ek = \frac{Ro}{Re} = \frac{\nu}{fR^2} ,
\ee
respectively, where $Re=UR/\nu>0$ is the Reynolds number. Therefore, although $Ek$ and $Ro$ can take both positive and negative values, they are either both positive or both negative.

Equations of motion \rf{eom2} are expressed in their dimensionless form as
\begin{subequations}
\label{eom3}
\begin{align}
Ro \frac{\dd \bm{u}}{\dd t} + \bm{e}_z \times \bm{u} &= - \bm{\nabla}_{\alpha} P - \frac{\rho}{\alpha^2} \bm{e}_z + Ek \mathcal{D} \bm{u} , \label{eom3-1} \\
Ro \frac{\dd \rho}{\dd t}  &= \frac{Ek}{Sc} \mathcal{D} \rho , \label{eom3-2} \\
\bm{\nabla} \cdot \bm{u} &= 0 , \label{eom3-3}
\end{align}
\end{subequations}
where $\dd/\dd t=\partial_t + (\bm{u} \cdot \bm{\nabla})$, $\mathcal{D}=\bm{\nabla}_h^2 + \alpha^{-2}\partial^2_z$ and $\bm{\nabla}_{\alpha} = (\partial_r , r^{-1}\partial_{\theta} , \alpha^{-2}\partial_z )^{T}$ is the modified gradient operator.

\section{Steady state}

The background flow is assumed to be purely azimuthal
\be{bfv}
\bm{U} = \left[ \bm{U}_h , U_z \right] = \left[ U_r , U_{\theta} , U_z \right] = \left[ 0 , r\Omega(r,z) , 0 \right] ,
\ee
where $\Omega(r,z)=(R/U)\tilde{\Omega}$ is the dimensionless angular velocity, Fig.~\ref{sketch}. Additionally, we assume the vortex profile to possess a Gaussian shape along both radial and axial directions
\be{av}
\Omega(r,z) = e^{- r^2 - z^2 }>0,
\ee
see Fig.~\ref{sketch}.
The profile \rf{av} represents a particular class of lenticular vortices, known as the Gaussian lenses. This model is adopted by the majority of theoretical studies of coherent isolated vortices because it fits both real oceanic Meddies and laboratory lenticular vortices reasonably well \cite{EBM2016,YSB2019,GC2003,GCB2004,EGCL2020,LB2021}.

Consider the equilibrium governed by the stationary and inviscid form of \rf{eom3-1} 
\begin{subequations}
\label{euler}
\begin{align}
\frac{\partial P}{\partial r} &= r \Omega \left( 1 + Ro \Omega \right) , \label{eul1} \\
\frac{\partial P}{\partial z} &= - \rho, \label{eul2}
\end{align}
\end{subequations}
where $\rho$ is the dimensionless version of the density profile \rf{rho}
\ba{rhodl}
\rho (r,z)  &=& \frac{g\alpha}{fU} - \left( \frac{g \alpha}{\rho_0 f U}\right)\rho_0\frac{ N^2}{g}Zz + \rho_A(r,z)\nn \\
&=&\frac{g\alpha}{fU} - \frac{\alpha^2 Ro}{F_h^2} z + \rho_A(r,z)\nn\\
&=&\frac{g\alpha}{fU} - \frac{Bu}{Ro}z + \rho_A(r,z)
\ea
and $Bu$ is the Burger number
\be{Bu}
Bu = \frac{\alpha^2 Ro^2}{F_h^2}.
\ee

Taking the radial derivative of expression \rf{eul2} and substituting expression \rf{eul1} in the result, we obtain {\color{black}{(in contrast to the thermal wind equation in\cite{M1970,LB2021})}} the \textit{gradient wind equation} \cite{EBM2016}  for the density profile \rf{rhodl} as
\be{gwe}
r\frac{\partial}{\partial z} \left[ \Omega \left( 1 + Ro \Omega \right) \right] = - \frac{\partial \rho_A}{\partial r}.
\ee
Making use of the angular velocity profile \rf{av} in  \rf{gwe} and computing the axial derivative yields
\be{gwet}
 2 rz \Omega \left( 1 + 2 Ro \Omega \right) = \frac{\partial \rho_A}{\partial r}.
\ee
Integrating \rf{gwet} over the radial coordinate returns an explicit expression for the density anomaly
\be{rhoa}
\rho_A(r,z) = - z \Omega \left( 1 + Ro \Omega \right).
\ee

Hence, the \textit{cyclogeostrophic balance} \cite{GZP2009,BGC2021,DV2003} between centrifugal, Coriolis and pressure forces yields
\be{rdiml}
\rho(r,z)=\frac{g\alpha}{fU} - z\frac{Bu}{Ro} - z \Omega \left( 1 + Ro \Omega \right).
\ee

Finally, we check that \rf{rdiml} satisfies an expression for the aspect ratio of a stratified vortex submerged in a stratified fluid  \cite{Au2012,HML2012}. Assume for simplicity that 
\be{}
\left. \frac{\partial \rho}{\partial z} \right|_{\substack[b]{r=0\\z=0}} = 0,
\ee
which corresponds to the limit $N_c\to 0$ ($N_c$ is the Brunt–V\"ais\"al\"a frequency associated with the vortex centre \cite{Au2012,HML2012,LB2021}) that eliminates a possibility for a gravitational instability of the vortex \cite{YSB2019}. Take into account that
$\rho=\rho_L+\rho_A$ and $\rho_L = (g\alpha)/(fU) - z Bu/Ro$. Then, with our profiles \rf{av} and \rf{rhoa}, implying $\Omega(0,0)=1$, we obtain 
\begin{align}
\left. \frac{\partial \rho}{\partial z} \right|_{\substack[b]{r=0\\z=0}} &= \left. \frac{\partial \rho_L}{\partial z} \right|_{\substack[b]{r=0\\z=0}} + \left. \frac{\partial \rho_A}{\partial z} \right|_{\substack[b]{r=0\\z=0}} \nn \\ &= - \frac{Bu}{Ro} - (Ro + 1) = 0.
\label{releg}
\end{align}
With the expressions \rf{Bu} and \rf{FhRo}, the relation \rf{releg} yields the aspect ratio of the Gaussian lens in an equilibrium
\be{asper}
\alpha^2 = -\frac{Ro ( 1 + Ro )}{N^2} f^2,
\ee
in full agreement with the general result \cite{Au2012,HML2012}.

\section{Linearized equations of motion}

We assume the background flow presented in the previous section to depart slightly from its original state, according to infinitesimal disturbances. Let thus introduce perturbations $(u_r',u_{\theta}',u_z',p',\rho')$ of velocity, pressure and density to perform a linear stability analysis of this hydrodynamical model.

Linearizing equations \rf{eom3} about the base state described in the previous section yields
\begin{subequations}
\label{leom}
\begin{align}
Ro \left( \frac{\dd}{\dd t} + \mathcal{U} \right) \bm{u}' + e_z \times \bm{u}' + \bm{\nabla}_{\alpha} p' + \frac{\rho'}{\alpha^2}  \bm{e}_z &=  Ek \mathcal{D} \bm{u}', \label{leom1} \\
Ro \left( \frac{\dd \rho'}{\dd t} + \mathcal{B}^{T} \bm{u}' \right) &= u_z' Bu + \frac{Ek}{Sc} \mathcal{D} \rho', \label{leom2} \\
\bm{\nabla} \cdot \bm{u}' &= 0, \label{leom3}
\end{align}
\end{subequations}
where $\dd/\dd t=\partial_t + (\bm{U} \cdot \nabla)$,  $\mathcal{U} = \bm{\nabla} \bm{U}$ and $\mathcal{B} = \bm{\nabla} \rho_A$ with
\begin{align}
\label{gradU}
\mathcal{U} = 
\begin{pmatrix}
0 & -\Omega & 0 \\
\Omega + r\partial_r \Omega & 0 & r \partial_z \Omega \\ 
0 & 0 & 0
\end{pmatrix} ,
\end{align}
and
\begin{align}
\label{gradrho}
\mathcal{B}  = 
\begin{pmatrix}
-z\dfrac{\partial\Omega}{\partial r} \left( 1 + 2Ro\Omega \right) \\
0 \\
- z \dfrac{\partial\Omega}{\partial z} \left( 1 + 2Ro\Omega \right) - \Omega \left( 1 + Ro\Omega \right) \\
\end{pmatrix} .
\end{align}

\section{Geometric optics approximation}

{\color{black}{Focusing on short-wavelength instabilities, we take advantage of the geometric optics approach \cite{KSF2014,KM2017,SM2019,Vidal2019} that provides a systematic procedure to finding asymptotic solutions of the linearized equations of motion \rf{leom} as an expansion in terms of a small formal parameter $\epsilon$, such that $0<\epsilon\ll 1$.}} Disturbances of the flow under such asymptotic expansion are written as \cite{KSF2014,KM2017}
\begin{subequations}
\label{goa}
\begin{align}
\bm{u}'(\bm{x},t,\epsilon) &= e^{i\frac{\phi(\bm{x},t)}{\epsilon}} \left[ \bm{u}^{(0)}(\bm{x},t) + \epsilon \bm{u}^{(1)}(\bm{x},t) \right] + \epsilon \bm{u}^{(r)}(\bm{x},t,\epsilon) , \\
p'(\bm{x},t,\epsilon) &= e^{i\frac{\phi(\bm{x},t)}{\epsilon}} \left[ p^{(0)}(\bm{x},t) + \epsilon p^{(1)}(\bm{x},t) \right] + \epsilon p^{(r)}(\bm{x},t,\epsilon) , \\
\rho'(\bm{x},t,\epsilon) &= e^{i\frac{\phi(\bm{x},t)}{\epsilon}} \left[ \rho^{(0)}(\bm{x},t) + \epsilon \rho^{(1)}(\bm{x},t) \right] + \epsilon \rho^{(r)}(\bm{x},t,\epsilon) , 
\end{align}
\end{subequations}
with $\phi$ being the phase of oscillations and $\bm{x}$ the vector of coordinates. We further assume the residual terms $[ \bm{u}^{(r)},p^{(r)},\rho^{(r)} ]$ to be uniformly bounded in $\epsilon$ \cite{KSF2014,KM2017}.

Substituting the series \rf{goa} in the isochoric condition \rf{leom3}, and retaining only terms of orders $\epsilon^{-1}$ and $\epsilon^{0}$, respectively, yields
\ba{}
    \bm{u}^{(0)} \cdot \bm{\nabla} \phi &= 0, \label{ic0} \\
    \bm{\nabla} \cdot \bm{u}^{(0)}+ i \bm{u}^{(1)} \cdot \bm{\nabla} \phi &= 0. \label{ic1}
\ea
Following the earlier works \cite{KSF2014,KM2017,K2017,K2021,SM2019,Vidal2019}, we assume that damping terms are quadratic in the small parameter $\epsilon$ and we therefore have $Ek=\epsilon^2\widetilde{Ek}$. Using a similar analysis as for the expanded incompressibility conditions \rf{ic0} and \rf{ic1}, we recover the Navier-Stokes equations \rf{leom1} along with the local conservation of density \rf{leom2} in terms of a linear system at order $\epsilon^{-1}$

\begin{widetext}
\begin{align}
\label{eoma1o}
    Ro \begin{pmatrix}
    \dfrac{\partial\phi}{\partial t} + \left( \bm{U} \cdot \bm{\nabla}\phi \right) & 0 \\
    0 & \dfrac{\partial\phi}{\partial t} + \left( \bm{U} \cdot \bm{\nabla}\phi \right)
    \end{pmatrix}
    \begin{pmatrix}
    \vphantom{\dfrac{\partial\phi}{\partial t}} \bm{u}^{(0)} \\
    \vphantom{\dfrac{\partial\phi}{\partial t}} \rho^{(0)}
    \end{pmatrix} = - \bm{\nabla}_{\alpha} \phi
    \begin{pmatrix}
    \vphantom{\dfrac{\partial\phi}{\partial t}} p^{(0)} \\
    \vphantom{\dfrac{\partial\phi}{\partial t}} 0
    \end{pmatrix} ,
\end{align}
\end{widetext}

\clearpage

and at order $\epsilon^0$


\begin{widetext}
\ba{eoma2o}
    i Ro
    \begin{pmatrix}
    \dfrac{\partial\phi}{\partial t} + \left( \bm{U} \cdot \bm{\nabla}\phi \right) & 0 \\
    0 & \dfrac{\partial\phi}{\partial t} + \left( \bm{U} \cdot \bm{\nabla}\phi \right)
    \end{pmatrix}
    \begin{pmatrix}
    \vphantom{\dfrac{\partial\phi}{\partial t}} \bm{u}^{(1)} \\
    \vphantom{\dfrac{\partial\phi}{\partial t}} \rho^{(1)}
    \end{pmatrix} &+& \nn \\
    \begin{pmatrix}
    Ro \left[ \dfrac{\partial}{\partial t} + \mathcal{U} + \bm{U} \cdot \bm{\nabla} \right] + \widetilde{Ek} \left( \bm{\nabla}_{\alpha} \phi \cdot \bm{\nabla} \phi \right) + \bm{e}_z \times  & \dfrac{\bm{e}_z}{\alpha^2} \\
    Ro \mathcal{B}^{T} - \bm{e}_z^{T} Bu & Ro \left[ \dfrac{\partial}{\partial t} + \bm{U} \cdot \bm{\nabla} \right] + \dfrac{\widetilde{Ek}}{Sc} \left( \bm{\nabla}_{\alpha} \phi \cdot \bm{\nabla} \phi \right)
    \end{pmatrix}
    \begin{pmatrix}
    \vphantom{\dfrac{\partial\phi}{\partial t}} \bm{u}^{(0)} \\
    \vphantom{\dfrac{\partial\phi}{\partial t}} \rho^{(0)}
    \end{pmatrix} &=& \nn \\
    - i\bm{\nabla}_{\alpha} \phi
    \begin{pmatrix}
    \vphantom{\dfrac{\partial\phi}{\partial t}} p^{(1)} \\
    \vphantom{\dfrac{\partial\phi}{\partial t}} 0
    \end{pmatrix}
    - \bm{\nabla}_{\alpha}
    \begin{pmatrix}
    \vphantom{\dfrac{\partial\phi}{\partial t}} p^{(0)} \\
    \vphantom{\dfrac{\partial\phi}{\partial t}} 0
    \end{pmatrix} &.&
\ea
\end{widetext}
Taking the dot product of the first equation in \rf{eoma1o} with $\bm{\nabla}\phi${\color{black}{, we obtain}}
\be{dp31}
{\color{black}{Ro \left( \bm{\nabla}\phi \cdot \bm{u}^{(0)} \right) \left[ \frac{\partial\phi}{\partial t} + \bm{U} \cdot \bm{\nabla}\phi  \right] = \left( \bm{\nabla}\phi \cdot \bm{\nabla}_{\alpha}\phi \right) p^{(0)} .}}
\ee
Applying the constraint \rf{ic0} {\color{black}{on \rf{dp31}}} yields \cite{KM2017}
\be{p0}
p^{(0)}=0.
\ee
Taking \rf{p0} into account in the linear system \rf{eoma1o}, while seeking for non-trivial solutions, we recover the Hamilton-Jacobi equation from the computation of its determinant \cite{KM2017}
\be{hj}
\frac{\partial\phi}{\partial t} + \bm{U} \cdot \bm{\nabla}\phi = 0.
\ee
For the rest of this section we assume that $\bm{\nabla}\phi = \bm{k}$ and $\bm{\nabla}_{\alpha}\phi = \bm{k}_{\alpha}$, with $\bm{k}=(k_r,k_\theta,k_z)^{T}$ and $\bm{k}_{\alpha}=(k_r,k_\theta,k_z/\alpha^2)^{T}$. From the application of the gradient operator $\bm{\nabla}$ on equation \rf{hj}, it yields the following eikonal equation \cite{KSF2014,KM2017,K2017,K2021}
\be{eek}
\frac{\dd \bm{k}}{\dd t} = - \mathcal{U}^{T} \bm{k} .
\ee

Taking relations \rf{p0} and \rf{hj} into account within \rf{eoma2o} results in the coupled equations
\ba{}
\left( Ro \left[ \frac{\dd}{\dd t} + \mathcal{U} \right] + Ek + \bm{e}_z \times \right) \bm{u}^{(0)} + \frac{\rho^{(0)}}{\alpha^2} \bm{e}_z &=& -i\bm{k}_{\alpha} p^{(1)} , \label{ee1} \\
\left( Ro \frac{\dd}{\dd t} + \frac{Ek}{Sc} \right) \rho^{(0)} + \left( Ro \mathcal{B}^{T} - \bm{e}_z^{T} Bu \right) \bm{u}^{(0)} &=& 0 , \label{ee2}
\ea
where $Ek=\widetilde{Ek} \vert \bm{k}_{\alpha}^{T}\bm{k} \vert$.

Taking the dot product of \rf{ee1} with $\bm{k}^{T}$ from the left, in view of \rf{ic0} we can isolate the first-order pressure term in the right-hand side and express it in terms of  zeroth-order terms
\be{p1}
p^{(1)} = \frac{i\bm{k}^{T}}{\bm{k}^{T}\bm{k}_{\alpha}} \left[ \left( Ro \left[ \frac{\dd}{\dd t} + \mathcal{U} \right] + \bm{e}_z \times \right) \bm{u}^{(0)} + \frac{\rho^{(0)}}{\alpha^2} \bm{e}_z \right] .
\ee
Differentiating \rf{ic0} yields \cite{KSF2014,KM2017}
\be{vik}
\frac{\dd}{\dd t} \left( \bm{k} \cdot \bm{u}^{(0)} \right) = \frac{\dd \bm{k} }{\dd t} \cdot \bm{u}^{(0)} + \bm{k} \cdot \frac{\dd\bm{u}^{(0)}}{\dd t} = 0. 
\ee

With the identity \rf{vik} the expression \rf{p1} becomes
\ba{p12}
p^{(1)} &=& \frac{i\bm{k}^{T}}{\bm{k}^{T}\bm{k}_{\alpha}} \left[ Ro \, \mathcal{U} \bm{u}^{(0)} + \bm{e}_z \times \bm{u}^{(0)} + \frac{\rho^{(0)}}{\alpha^2} \bm{e}_z \right] \nn\\
&-& \frac{iRo}{\bm{k}^{T}\bm{k}_{\alpha}} \frac{\dd\bm{k}}{\dd t} \cdot \bm{u}^{(0)}.
\ea
Re-writing \rf{p12} by means of the phase equation \rf{eek}, we further obtain
\be{p1k}
p^{(1)} = \frac{i \bm{k}^{T}}{\beta^2} \left( \bm{e}_z \times \bm{u}^{(0)} + \frac{\rho^{(0)}}{\alpha^2} \bm{e}_z \right) + 2 i Ro\frac{\bm{k}^{T}\mathcal{U}}{\beta^2} \bm{u}^{(0)} ,
\ee
where $\beta^2 = \bm{k}^{T}\bm{k}_{\alpha} = k_r^2 + k_{\theta}^2 + k_z^2/\alpha^2$.

Inserting expression \rf{p1k} in \rf{ee1} yields the transport equations
\ba{te}
Ro \frac{\dd \bm{u}^{(0)}}{\dd t} &=& - Ek \bm{u}^{(0)} - Ro \left( \mathcal{I} - 2\frac{\mathcal{K}}{\beta^2} \right) \mathcal{U} \bm{u}^{(0)} \nn \\  
&-& \left( \mathcal{I} - \frac{\mathcal{K}}{\beta^2} \right) \bm{e}_z \times \bm{u}^{(0)} - \left( \mathcal{I} - \frac{\mathcal{K}}{\beta^2} \right) \bm{e}_z \frac{\rho^{(0)}}{\alpha^2}  , \nn \\
Ro \frac{\dd \rho^{(0)}}{\dd t} &=& - \frac{Ek}{Sc} \rho^{(0)} - \left( \mathcal{B}^{T} Ro - \bm{e}_z^{T} Bu \right) \bm{u}^{(0)} ,
\ea
where  $\mathcal{I}$ is the identity matrix and $\mathcal{K} = \bm{k}_{\alpha} \bm{k}^{T}$.

From the eikonal equation \rf{eek} we deduce that $k_r=k_z=const$ and $k_\theta=0$ due to expression \rf{gradU} \cite{KSF2014,KM2017,K2017,K2021}. Introducing the scaled wavenumbers $q_r=k_r/\beta$ and $q_z=k_z/\beta$, we find  $q_r = \sqrt{1 - q_z^2/\alpha^2}$. This allows us to write $\mathcal{K} = \bm{q}_{\alpha} \bm{q}^{T}$, where $\bm{q}=(q_r,0,q_z){\color{black}{^{T}}}$ and $\bm{q}_{\alpha}=(q_r,0,q_z/\alpha^2){\color{black}{^{T}}}${\color{black}{, such that}}
\be{}
{\color{black}{
\mathcal{K} = 
\begin{pmatrix}
q_r^2 & 0 & q_r q_z \\
0 & 0 & 0 \\ 
q_r q_z / \alpha^2 & 0 & q_z^2/\alpha^2
\end{pmatrix} .
}}
\ee

In the new notation the amplitude transport equations \rf{te} for the perturbed velocity and density fields take the following explicit form
\begin{widetext}
\begin{subequations}
\label{ate}
\begin{align}
\left[ Ro \left( \frac{\partial}{\partial t} + \Omega\frac{\partial}{\partial\theta} \right) + Ek \right] u_r^{(0)} - \frac{q_z^2}{\alpha^2} \left( 1 + 2 Ro \Omega \right) u_{\theta}^{(0)}  - \frac{q_z q_r}{\alpha^2} \rho^{(0)} &= 0 , \label{ate1} \\
\left[ Ro \left( \frac{\partial}{\partial t} + \Omega\frac{\partial}{\partial\theta} \right) + Ek \right] u_{\theta}^{(0)} + \left[ 1 + Ro \left( 2 \Omega + r\frac{\partial\Omega}{\partial r} \right) \right] u_r^{(0)} + r Ro \frac{\partial\Omega}{\partial z} u_z^{(0)} &= 0 , \label{ate2} \\
\left[ Ro \left( \frac{\partial}{\partial t} + \Omega\frac{\partial}{\partial\theta} \right) + Ek \right] u_z^{(0)} + \frac{q_z q_r}{\alpha^2} \left( 1 + 2 Ro \Omega \right) u_{\theta}^{(0)} + \frac{q_r^2}{\alpha^2} \rho^{(0)} &= 0 , \label{ate3} \\
\left[ Ro \left( \frac{\partial}{\partial t} + \Omega\frac{\partial}{\partial\theta} \right) + \frac{Ek}{Sc} \right] \rho^{(0)} - u_z^{(0)} Bu + Ro \left( \frac{\partial \rho_A}{\partial r} u_r^{(0)} + \frac{\partial \rho_A}{\partial z} u_z^{(0)} \right)  &= 0 , \label{ate4}
\end{align}
\end{subequations}
\end{widetext}

Observing that \rf{ate1} and \rf{ate3} coincide under the linear transformation $u_z^{(0)}= -(q_r/q_z) u_r^{(0)}$ we can eliminate the variable $u_z^{(0)}$ and thus reduce the number of equations in the system \rf{ate} to three \cite{KM2017}, with respect to $u_r^{(0)}$, $u_{\theta}^{(0)}$, and $\rho^{(0)}$ only.

\section{Dispersion relation}

Introducing in the reduced system \rf{ate} the complex growth rate $\lambda$ and the azimuthal wavenumber $m$ from the ansatz \cite{KM2017,KSF2014,K2017}
\be{}
\left[ u_r^{(0)}, u_{\theta}^{(0)}, \rho^{(0)} \right] = \left[ \hat{u}_r^{(0)}, \hat{u}_{\theta}^{(0)}, \hat{\rho}^{(0)} \right] \exp{\left( \lambda t + i m\theta\right)} ,
\ee
yields a linear eigenvalue problem $\mathcal{H}\xi=\hat{\lambda}\xi$, where $\xi=\left( \hat{u}_r^{(0)}, \hat{u}_{\theta}^{(0)}, \hat{\rho}^{(0)} \right)^{T}$, $\hat{\lambda}=Ro\left(\lambda +im\Omega\right)+Ek$, and $\mathcal{H}$ is a $3\times 3$ matrix
\be{H}
\mathcal{H} = 
\begin{pmatrix}
  0 & \dfrac{q_z^2}{\alpha^2} \dfrac{2j}{r^2} & \dfrac{q_r q_z }{\alpha^2} \\
  -\dfrac{r^2 \left( q_z \kappa_r^2 - q_r \kappa_z^2 \right)}{2 q_z j}  & 0 & 0\\
  - \dfrac{ Ro \left( q_z \partial_r \rho_A - q_r \partial_z \rho_A \right) + q_r Bu }{q_z}  & 0 & Ek \dfrac{Sc - 1}{Sc}
\end{pmatrix} , 
\ee
where $\partial_r\rho_A$ and $\partial_z\rho_A$ are given by \rf{gradrho}, $j(r,z)$ is the angular momentum per unit mass 
\be{j}
j = \frac{r^2}{2} \left( 1 + 2 Ro \Omega \right) ,
\ee
$\kappa_r$ is the epicyclic frequency and $\kappa_z$ is the  frequency of vertical oscillations   \cite{LU2019}
\be{krkz}
\kappa_r^2 = r^{-3} \frac{\partial j^2}{\partial r}, \quad \kappa_z^2 = r^{-3} \frac{\partial j^2}{\partial z}.
\ee

The dispersion relation $\mathcal{D}(\hat{\lambda})$ of the system is obtained from
\be{dr}
\mathcal{D}(\hat{\lambda}) = \det \left( \mathcal{H} - \hat{\lambda}\mathcal{I} \right) ,
\ee
and is a third-order polynomial in $\hat{\lambda}$ 
\be{drfull2}
\mathcal{D}(\hat{\lambda}) = \hat{\lambda}^3 + Ek \frac{1-Sc}{Sc} \hat{\lambda}^2 + \left( \gamma_1 + \gamma_2 \right) \hat{\lambda} + Ek \frac{1-Sc}{Sc} \gamma_1
\ee
where
\ba{gamma}
\gamma_1 &=& \frac{q_z}{\alpha^2} \left( q_z \kappa_r^2 - q_r  \kappa_z^2 \right), \nn \\
\gamma_2 &=& \frac{q_r}{\alpha^2} \left[ Ro \left( q_z \frac{\partial\rho_A}{\partial r} - q_r \frac{\partial\rho_A}{\partial z} \right) + q_r Bu \right] .
\ea

It is worth mentioning that similar dispersion relations of third order were obtained earlier by McIntyre\cite{M1970}, who studied a baroclinic circular vortex in the presence of viscosity and a temperature gradient, and Singh and Mathur \cite{SM2019} who studied a barotropic columnar vortex in a stratified ambient fluid in the non-rotating frame. In both of these works, the authors restricted their analyses to axisymmetric $(m=0)$ instabilities only. 

Therefore, dispersion relation \rf{drfull2} with the coefficients \rf{gamma} substantially generalizes those of the previous works as it takes into account rotation of the frame, azimuthal wavenumber $m$, diffusion of mass and momentum, both $r$- and $z$-dependence of the vortex angular velocity $\Omega$ via the Gaussian profile \rf{av}, radial and axial stratification of the vortex, its aspect ratio, and axial stratification of the ambient fluid. This implies that a shear parameter $\gamma_1$ related to differential rotation induced by the vortex and a buoyancy parameter $\gamma_2$ related to the density stratification of the ambient fluid influenced by the vortex, can take both positive and negative values.

\begin{figure*}[t!]
\centering
    \begin{subfigure}{0.33\textwidth}
    \includegraphics*[width=\textwidth]{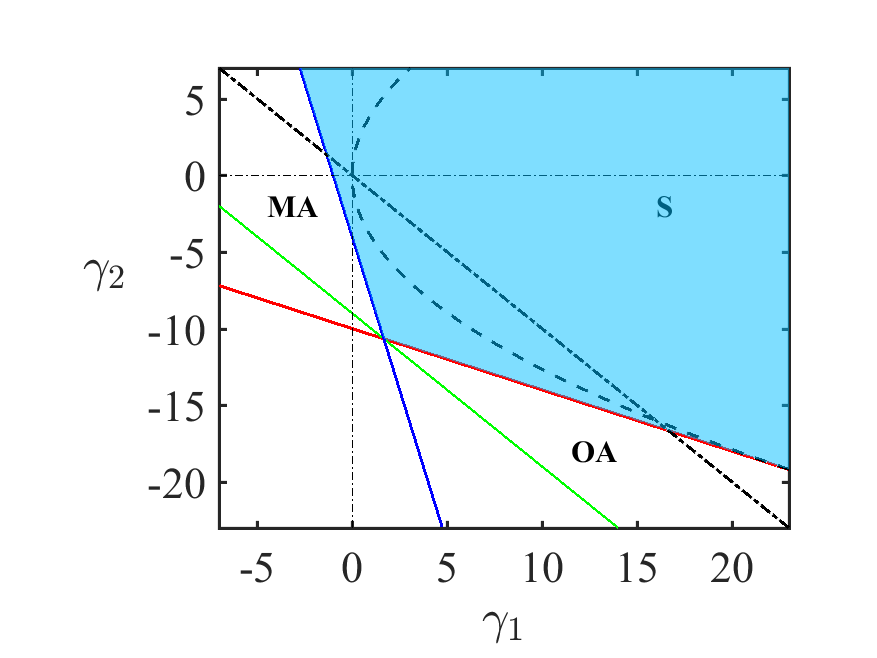}
    \caption{}\label{}
    \end{subfigure}
    \begin{subfigure}{0.33\textwidth}
    \includegraphics*[width=\textwidth]{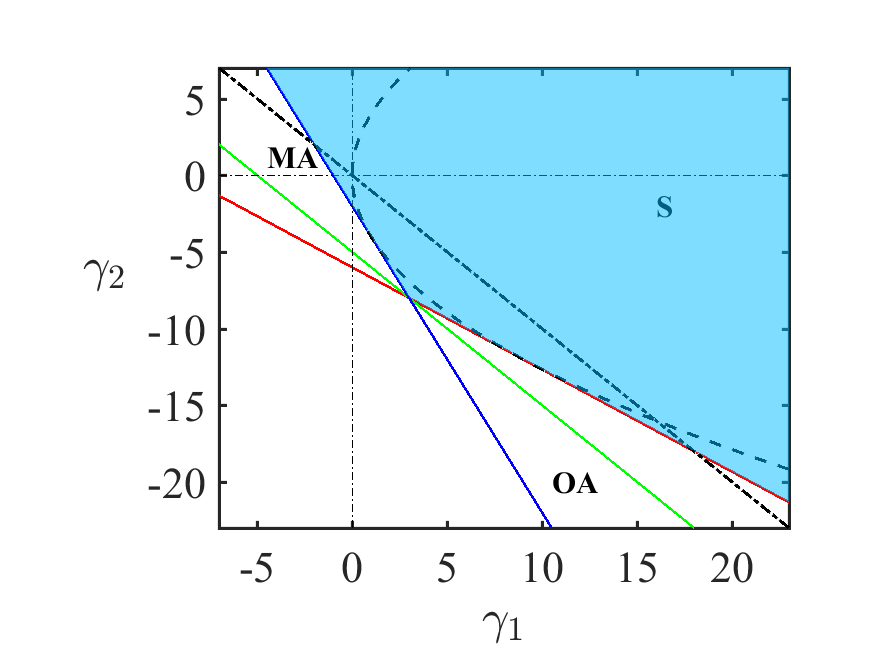}
    \caption{}\label{}
    \end{subfigure}
    \begin{subfigure}{0.33\textwidth}
    \includegraphics*[width=\textwidth]{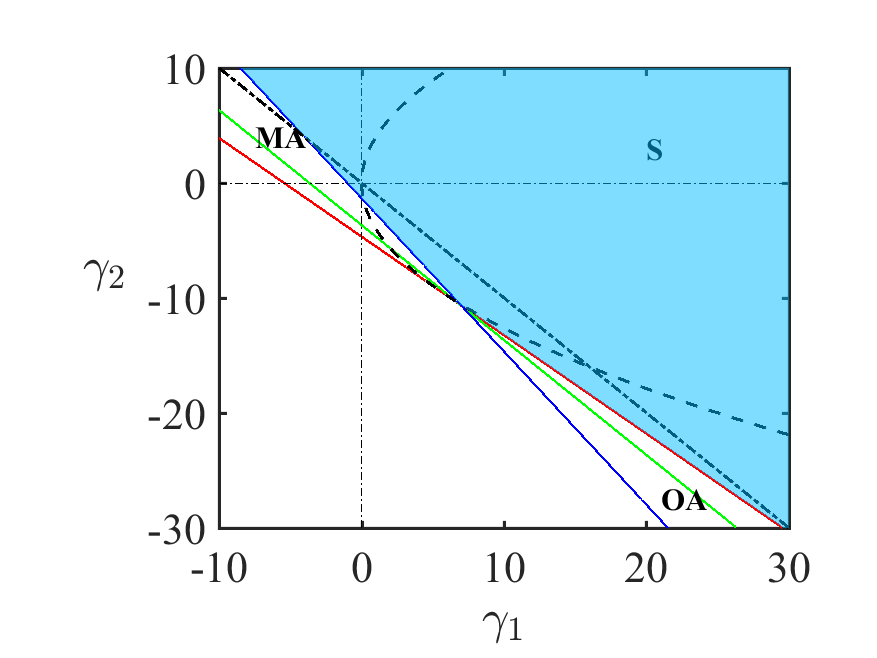}
    \caption{}\label{}
    \end{subfigure}\\
    \begin{subfigure}{0.33\textwidth}
    \includegraphics*[width=\textwidth]{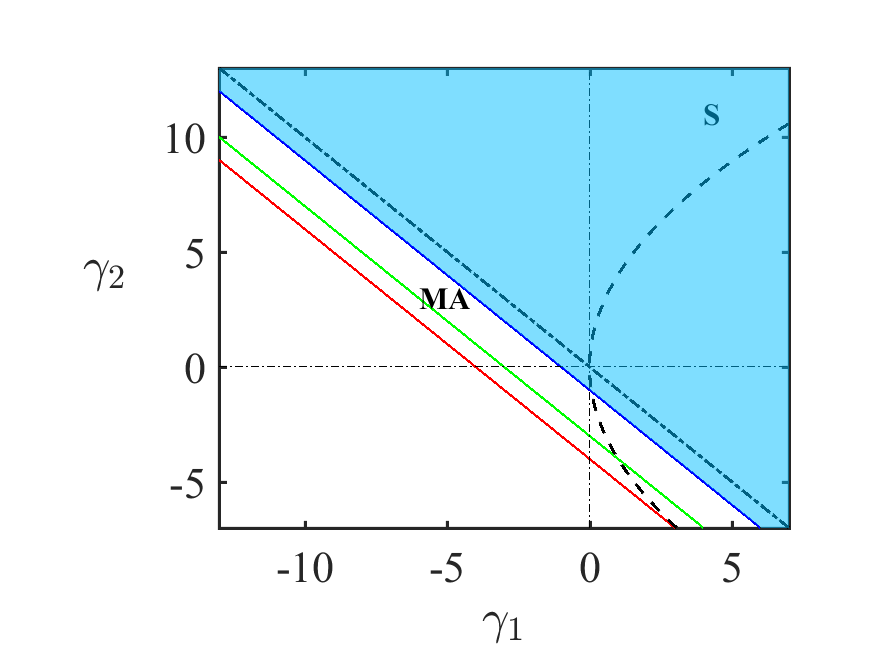}
    \caption{}\label{}
    \end{subfigure}
    \begin{subfigure}{0.33\textwidth}
    \includegraphics*[width=\textwidth]{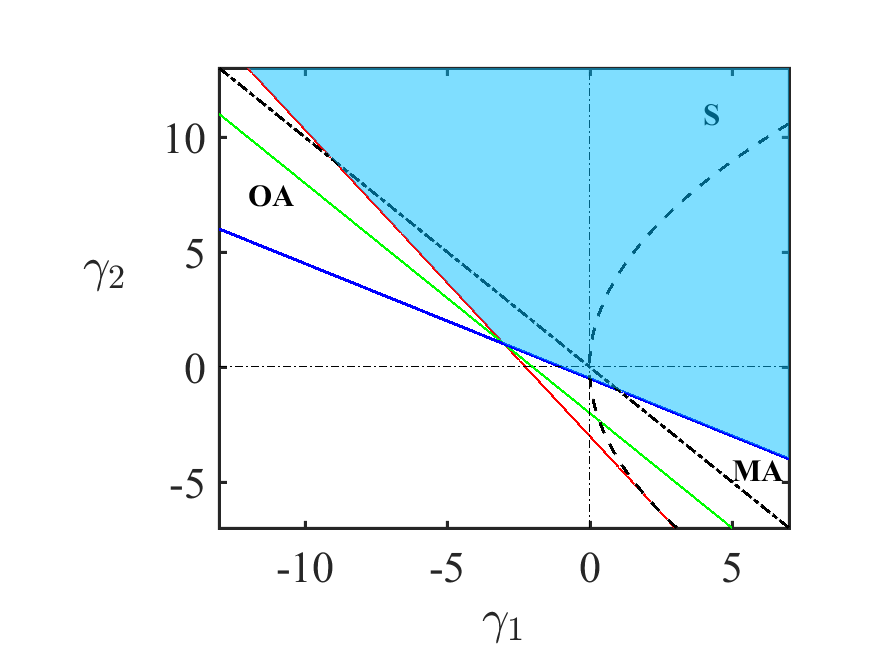}
    \caption{}\label{}
    \end{subfigure}
    \begin{subfigure}{0.33\textwidth}
    \includegraphics*[width=\textwidth]{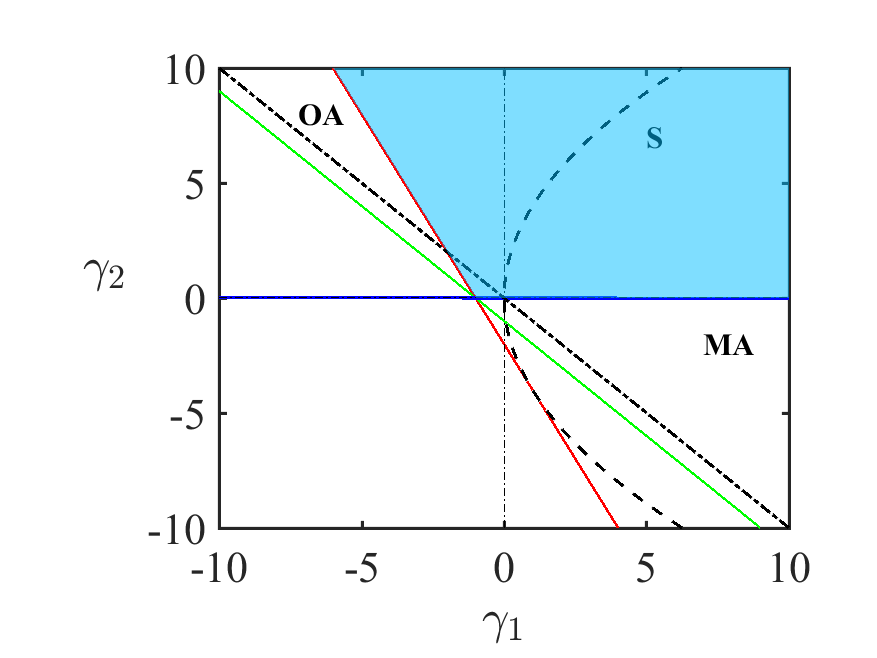}
    \caption{}\label{}
    \end{subfigure}

  \caption{Stability maps with codimension-2 points \rf{codim2} on the neutral stability curve for $Ek=1$ and (a) $Sc=0.25$, (b) $Sc=0.5$, (c) $Sc=0.75$, (d) $Sc=1$, (e) $Sc=2$, (f) $Sc\rightarrow +\infty$. The blue solid line stands for the boundary of the domain of {\color{black}{monotonic}} axisymmetric {\color{black}{(MA)}} instability \rf{axisym3}, the red solid line for that of the oscillatory axisymmetric (OA) instability \rf{axisym1}, S stands for the stability domain. The dashed line is the envelope \rf{envelop} and the dot-dashed line is the neutral stability boundary for the diffusionless system \rf{g1pg2}. The green solid line corresponds to the condition \rf{axisym2}. \label{fig:codim2g1g2}}
\end{figure*}

\subsection{Diffusionless and $Sc=1$ cases}

Notice that at $Ek=0$, as well as at $Sc=1$, dispersion relation \rf{drfull2} factorizes into a product of quadratic and linear in $\hat{\lambda}$ polynomials and thus can be solved explicitly.

In these cases, the eigenvalues governing the centrifugal instability are recovered as, cf. \cite{BG2005}
\be{eiginv}
\lambda^{\pm} = - \frac{im\Omega}{Ro} \pm \frac{1}{Ro} \sqrt{-\left( \gamma_1 + \gamma_2 \right)} ,
\ee
which reads the instability condition 
\be{g1pg2}
\gamma_1 + \gamma_2  <0.
\ee

\subsubsection{Centrifugal instability of a barotropic circular vortex}
In the particular case of purely transverse perturbations ($q_r\to 0$), the eigenvalues \rf{eiginv} are 
\be{}
\lambda^{\pm} = - \frac{im\Omega}{Ro} \pm \frac{q_z}{\alpha Ro} \sqrt{-\kappa_r^2} ,
\ee
and therefore, yield an instability when $\kappa_r^2<0$. Notice that in the dimensional variables and parameters
$$\kappa_r^2=f^{-2}(\partial_{\tilde r} \tilde{u}_{z}+\tilde{u}_{z}/\tilde{r}+f)(2\tilde{u}_{z}/\tilde{r}+f)$$
is nothing else but the generalized Rayleigh discriminant for a barotropic circular vortex \cite{YSB2019}. The inequality $\kappa_r^2<0$ is thus
the well-known criterion for the centrifugal instability of columnar vortices\cite{M1992,YSB2019}.

\subsubsection{Connection to Acheson and Gibbons\cite{AG78}}

Let us compare our criterion for centrifugal instability \rf{g1pg2} with the results derived previously by Acheson and Gibbons in the study of a magnetic and differentially rotating star \cite{AG78}.

For this purpose, we first present their axisymmetric and diffusionless dispersion relation (without magnetic field) in its original form
\be{agdr}
\gamma \frac{s^2}{n^2} \omega^2 = \gamma r \Omega^2 \frac{\partial R}{\partial h} + G \frac{\partial E}{\partial h} ,
\ee
where, in our notations, $n=k_z$ is the axial wavenumber, $s=\vert\bm{k}\vert$ is the norm of the wave vector, $\sigma=i\lambda$ is an eigenfrequency, $\omega=\sigma-m\Omega$ is the Doppler-shifted eigenfrequency,  $R=\ln{r^4\Omega^2}$ is the {\color{black}{logarithm of}} squared angular momentum, $E=\ln{p\rho^{-\gamma}}$ is a measure of entropy, $G=g_r-(k_r/k_z)g_z$ is a function containing gravitational effects and $\gamma$ is the heat capacity ratio. The derivative operator in \rf{agdr} is further defined as $\partial/\partial h=\partial/\partial r - (k_r/k_z)\partial/\partial z$ \cite{AG78}.

Multiplying both sides of \rf{agdr} by $n^2/s^2$ and introducing the wavenumbers $q_r=k_r/s$ and $q_z=k_z/s$, we first recover
\be{cv1}
\left( \lambda + im\Omega \right)^2 = - q_z^2 \left[  \frac{\tilde{j}^2}{r^3} \frac{\partial_h \tilde{j}^2}{\tilde{j}^2} + \frac{G}{\gamma} \partial_h (p\rho^{-\gamma}) \right] ,
\ee
where $\tilde{j}=r^2 \Omega$ is a simplified version of the angular momentum \rf{j} without the influence of the Coriolis force.

If the gravity is directed along the axial $z$-coordinate only (as it is in our setting), then using the correspondence $(g_r,g_z)=(0,\rho)$ within the function $G$ in \rf{cv1} yields
\be{cv2}
\left( \lambda + im\Omega \right)^2 = - q_z^2 \left[  \frac{1}{r^3} \partial_h \tilde{j}^2 - \frac{q_r\rho}{q_z\gamma} \left(  \frac{\partial_h p}{p} + \gamma \frac{\partial_h \rho}{\rho}  \right)  \right].
\ee

As a consequence of the Newton-Laplace equation, the specific heat capacity ratio is related to the speed of sound $c_s$ via the expression $c_s^2=\gamma p/\rho$ and hence, tends to infinity in the case of incompressible flows (as it is in our case). Taking this limit in \rf{cv2}, we obtain
\be{}
\left( \lambda + im\Omega \right)^2 = - q_z \left[ \left( q_z \tilde{\kappa}_r^2 - q_r \tilde{\kappa}_z^2 \right) + \frac{q_r}{q_z} \left( q_z \partial_r\rho - q_r \partial_z\rho \right) \right] ,
\ee
with $\tilde{\kappa}_z=r^{-3}\partial_z\tilde{j}^2$ and $\tilde{\kappa}_r=r^{-3}\partial_r\tilde{j}^2$. Finally, the inviscid eigenfrequency of the work \cite{AG78} takes the form
\be{eigag}
\lambda^{\pm} = - im\Omega \pm \sqrt{- \left[ q_z \left( q_z \tilde{\kappa}_r^2 - q_r \tilde{\kappa}_z^2 \right) + q_r \left( q_z \partial_r\rho - q_r \partial_z\rho \right) \right]}.
\ee
Notice that the radicand in \rf{eigag} has the same structure as our expressions \rf{gamma} and \rf{eiginv}, with the difference only in the factors $\alpha^{-2}$ and $Ro$ and in the term containing the Burger number.

\subsection{Particular cases when either $\gamma_1=0$ or $\gamma_2=0$}

In these two particular cases the polynomial \rf{drfull2} factorizes, which allows us to find its roots explicitly. 

For $\gamma_1=0$ the roots are
\ba{g10}
\lambda_{1,2} &=&-im\Omega-\frac{Ek}{2 Ro}\left\{1 + \frac{1}{Sc} \pm \sqrt{\frac{(Sc - 1)^2}{Sc^2}  - \frac{4\gamma_2}{Ek^2}}\right\}, \nn\\
\lambda_3&=&-im\Omega-\frac{Ek}{Ro}. 
\ea
Recalling that $Ek/Ro=1/Re>0 $, we see that in the limit $Sc \rightarrow +\infty$ the vortex is stable if $\gamma_2\ge0$, see  Fig.~\ref{fig:codim2g1g2}(f) and Fig.~\ref{fig:caustic}. In general, the condition for stability at $\gamma_1=0$ reads

\be{staco}
\gamma_2 \ge -\frac{Ek^2}{Sc}.
\ee

In the particular case when $\gamma_2=0$ we have
\ba{g20}
\lambda_{1,2} &=&-im\Omega-\frac{Ek}{Ro} \pm \frac{\sqrt{-\gamma_1}}{Ro}, \nn\\
\lambda_3&=&-im\Omega-\frac{Ek}{Ro Sc}.
\ea

According to \rf{g20}, in the diffusionless case $(Ek=0)$ the vortex is stable regardless of the sign of $Ro$, if and only if $\gamma_1\ge 0$, which is similar to the generalized Rayleigh criterion $\kappa_r^2-\kappa_z^2\ge 0$ described in the literature \cite{YSB2019}, since $\gamma_2=0$ corresponds to the radial and axial density gradients compensating each other. When diffusivities of mass and momentum are taken into account, then with any $Sc > 0$ such vortices remain stable. Unstable diffusionless vortices ($\gamma_1<0$)  can be stabilized for any $Sc > 0$, if $|\gamma_1|<Ek^2$. This is consistent with the results of McIntyre \cite{M1970} and Singh and Mathur \cite{SM2019}, see Fig.~\ref{fig:codim2g1g2}.

\section{General stability analysis}

\subsection{Bilharz algebraic criterion}

Written with respect to $\lambda$ the polynomial \rf{drfull2} has complex coefficients.
Bilharz algebraic criterion \cite{B1944,K2021} guarantees that all the roots of a complex polynomial of the form $p(\lambda) = (a_0 + ib_0) \lambda^3 + (a_1 + ib_1) \lambda^2 + (a_2 + ib_2) \lambda + (a_3+ib_3)$ lie in the open left half of the complex $\lambda$-plane if and only if {\color{black}{determinants of three even-order submatrices on the main diagonal (counting from the upper left corner)}} of the following Bilharz matrix,
\be{bilmat}
\mathcal{B} = 
\begin{pmatrix}
a_3 & -b_3 & 0 & 0 & 0 & 0 \\
b_2 & a_2 & a_3 & -b_3 & 0 & 0 \\
-a_1 & b_1 & b_2 & a_2 & a_3 & -b_3 \\
-b_0 & -a_0 & -a_1 & b_1 & b_2 & a_2 \\
0 & 0 & -b_0 & -a_0 & -a_1 & b_1 \\
0 & 0 & 0 & 0 & -b_0 & -a_0 
\end{pmatrix} ,
\ee
are strictly positive. In view of $Ek/Ro=1/Re>0$, being applied to polynomial \rf{drfull2}, the Bilharz criterion yields
\begin{widetext}
\begin{align}
\left\{m^4\Omega^4 Ro^4 Sc (2Sc + 1) + 6m^2\Omega^2Ro^2Sc(Ek^2 + Sc\gamma_2 + \gamma_1) + (Ek^2 + Sc\gamma_2 + \gamma_1)\left[(Sc + 2)Ek^2 + Sc(\gamma_1 + \gamma_2)\right]\right\}&\nn\\ \times \left[2(Sc + 1)^2Ek^2 + Sc(2Sc\gamma_1 + Sc\gamma_2 + \gamma_2)\right] &> 0,\label{bilh1}\\
m^4\Omega^4 Ro^4 Sc (2Sc + 1) + m^2Ro^2\Omega^2\left[2(Sc^2 + Sc + 1)Ek^2 + Sc(2\gamma_1 - \gamma_2)(Sc - 1)\right]& \nn\\ + (Ek^2 + Sc\gamma_2 + \gamma_1)\left[(Sc + 2)Ek^2 + Sc(\gamma_1 + \gamma_2)\right] &> 0,\label{bilh2} \\
Sc(Ek^2 + Sc\gamma_2 + \gamma_1) &> 0. \label{bilh3}
\end{align}
\end{widetext}
\begin{figure*}[t!]
\centering
   \begin{subfigure}{0.33\textwidth}
    \includegraphics*[width=\textwidth]{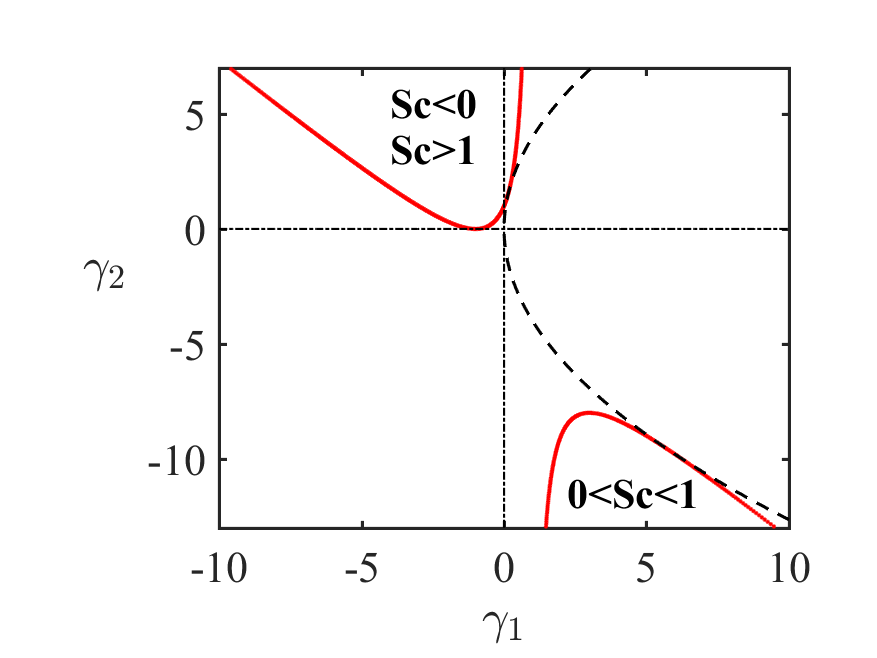}
    \caption{}\label{}
    \end{subfigure}
    \begin{subfigure}{0.33\textwidth}
    \includegraphics*[width=\textwidth]{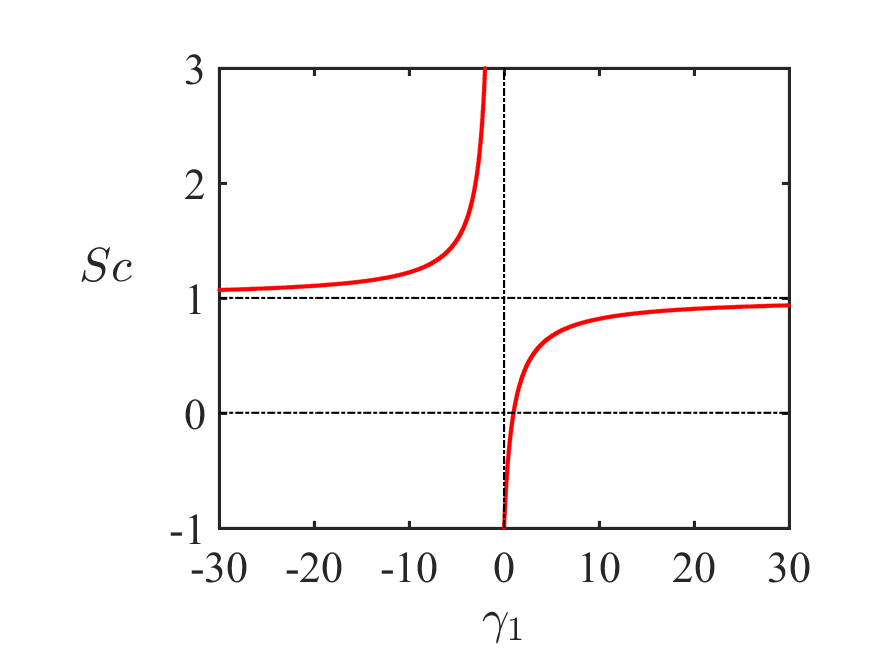}
    \caption{}\label{}
    \end{subfigure}
    \begin{subfigure}{0.33\textwidth}
    \includegraphics*[width=\textwidth]{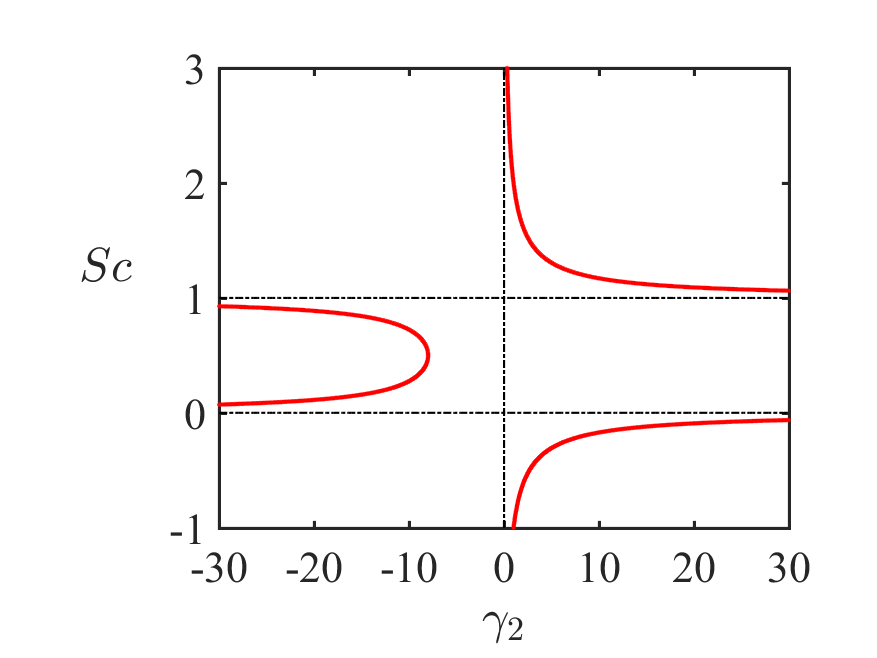}
    \caption{}\label{}
    \end{subfigure}

  \caption{For $Ek=1$ projections of the loci of the codimension-2 points onto {\color{black}{(a) $(\gamma_1,\gamma_2)$ plane, (b) $(\gamma_1,Sc)$ plane and (c) $(\gamma_2,Sc)$ plane,}} given by equations \rf{codim2}  and \rf{pg1g2}. The dashed curve is the envelope \rf{envelop}. \label{fig:loci}}
\end{figure*}
In the following, we will use the inequalities \rf{bilh1}--\rf{bilh3} to examine stability and instabilities of the lenticular vortex in the presence of differential diffusion of mass and momentum.

\subsection{{\color{black}{Monotonic}} and oscillatory axisymmetric instabilities }
%
%
%

\subsubsection{A codimension-2 point on the neutral stability line}

Setting $m=0$ in \rf{bilh1}--\rf{bilh3} we find that for $Sc>0$
the base flow is stable if and only if the following three inequalities are fulfilled simultaneously
\ba{axisym}
2(Sc + 1)^2Ek^2 + Sc(2Sc\gamma_1 + \gamma_2(Sc + 1))&>&0 , \label{axisym1}\\
(Sc + 2)Ek^2 + Sc(\gamma_1 + \gamma_2)   &>&0 , \label{axisym2}\\
Ek^2 + Sc\gamma_2 + \gamma_1 &>& 0 . \label{axisym3}
\ea

Although $Sc<0$ might not look physically meaningful, we mention, for completeness, that in this case the inequality \rf{axisym1} remains the same whereas the inequalities \rf{axisym2} and \rf{axisym3} are reversed.  It is worth to notice that continuation of stability diagrams to negative values of dissipation parameters can help in uncovering instability mechanisms \cite{K2017,K2021}.

The expressions in \rf{axisym1}--\rf{axisym3} are linear in $\gamma_1$ and $\gamma_2$ which makes it convenient to represent the criteria in the $(\gamma_1,\gamma_2)$-plane \cite{SM2019}, where the corresponding stability domain will be given by the intersection of the half-planes \rf{axisym1}--\rf{axisym3}, see Fig.~\ref{fig:codim2g1g2}.

 Equating to zero the left-hand sides of the expressions \rf{axisym1}--\rf{axisym3} and then solving the resulting equations with respect to $\gamma_1$ and $\gamma_2$, we find that all the three straight lines intersect at one and the same point with the coordinates, cf. \cite{SM2019}
\be{codim2}
\gamma_1 = Ek^2\frac{1 + Sc}{1 - Sc}, \quad \gamma_2 = \frac{2Ek^2}{Sc(Sc - 1)}.
\ee
%
At a given value of $Ek$ equations \rf{codim2} define a spatial curve in the $(\gamma_1,\gamma_2, Sc)$-space, which projections are shown in Fig.~\ref{fig:loci}. In particular, the projection onto the $(\gamma_1,\gamma_2)$-plane is 
\be{pg1g2}
\gamma_2(Ek^2 - \gamma_1) - (Ek^2 + \gamma_1)^2=0.
\ee

At the common point \rf{codim2} the slopes $\dd \gamma_2/\dd \gamma_1$ of straight lines \rf{axisym1}, \rf{axisym2}, and \rf{axisym3}  are, respectively,
\ba{slopes}
\sigma_1=-\frac{2Sc}{Sc + 1}, \quad \sigma_2=-1, \quad \sigma_3=-\frac{1}{Sc}.
\ea
Notice the following relationships between the slopes:
\ba{slor}
\left.
\begin{array}{r}
  -1 \ge \sigma_1 > -2 \\
    \sigma_2=-1 \\
-1  \le \sigma_3 < 0
\end{array}
\right\} \quad &{\rm if}& \quad 1 \le Sc < +\infty,\nn\\
\left.
\begin{array}{r}
  0 > \sigma_1 > -1 \\
    \sigma_2=-1 \\
-\infty  <\sigma_3 < -1
\end{array}
\right\} \quad &{\rm if}& \quad 0< Sc < 1.
\ea
\begin{figure*}[t!]
\centering
    \begin{subfigure}{0.32\textwidth}
    \includegraphics*[width=\textwidth]{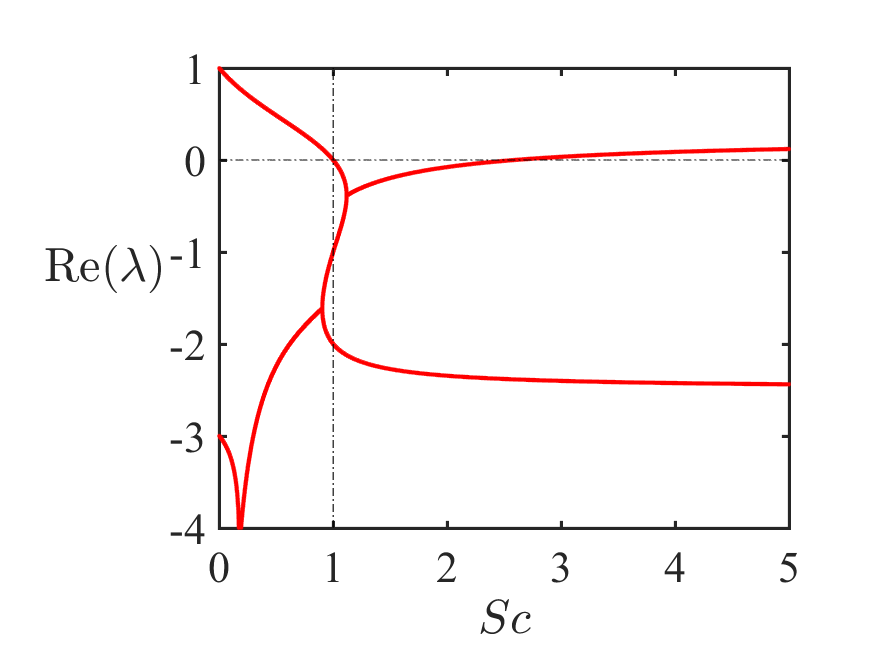}
    \caption{}\label{}
    \end{subfigure}
    \begin{subfigure}{0.32\textwidth}
    \includegraphics*[width=\textwidth]{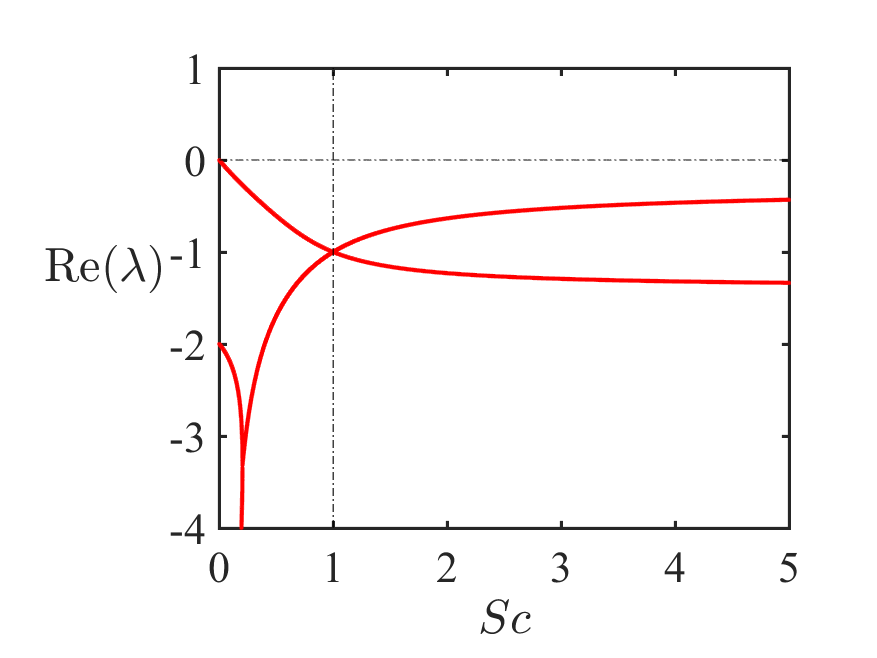}
    \caption{}\label{}
    \end{subfigure}
    \begin{subfigure}{0.32\textwidth}
    \includegraphics*[width=\textwidth]{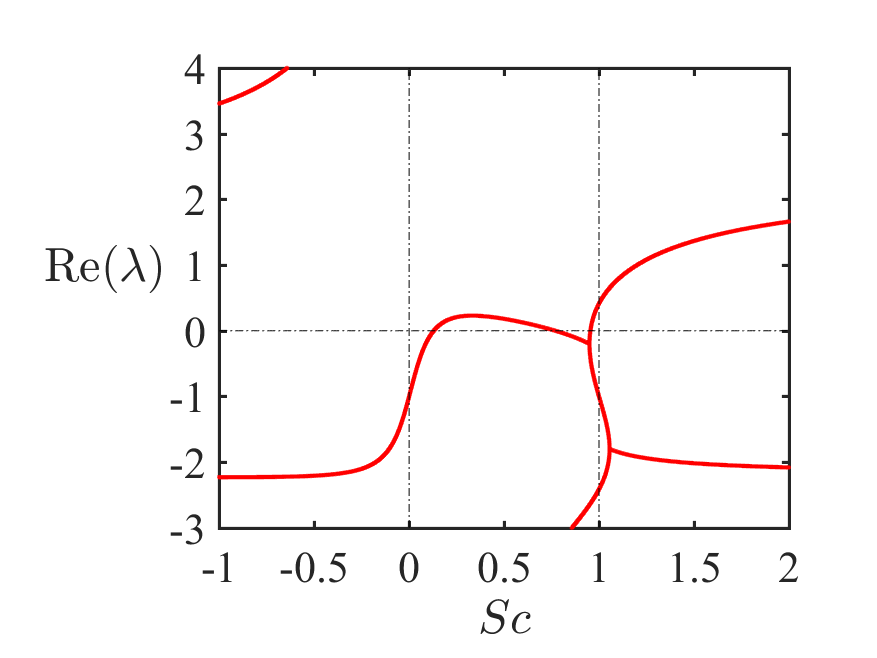}
    \caption{}\label{}
    \end{subfigure}\\
    \begin{subfigure}{0.32\textwidth}
    \includegraphics*[width=\textwidth]{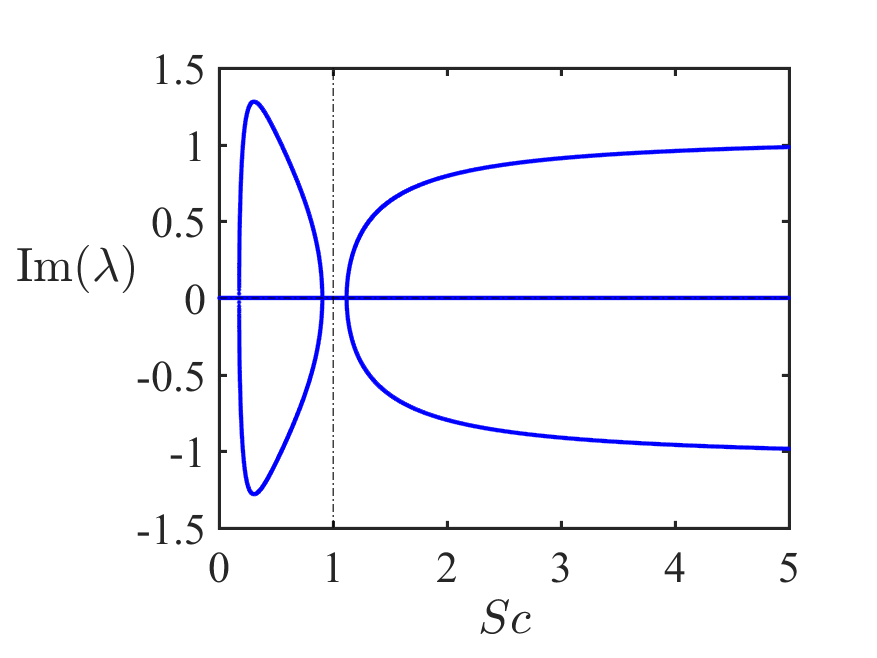}
    \caption{}\label{}
    \end{subfigure}
    \begin{subfigure}{0.32\textwidth}
    \includegraphics*[width=\textwidth]{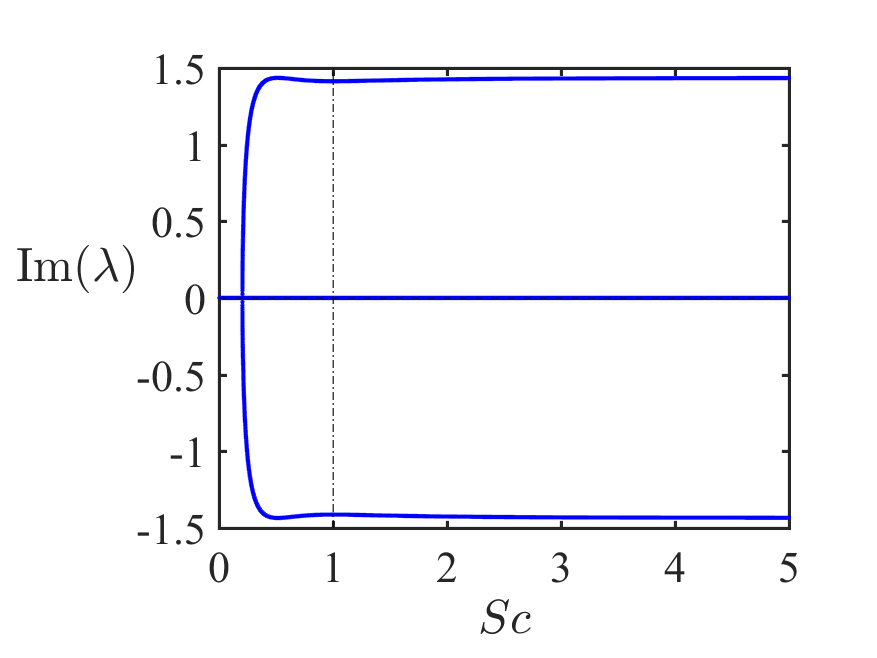}
    \caption{}\label{}
    \end{subfigure}
    \begin{subfigure}{0.32\textwidth}
    \includegraphics*[width=\textwidth]{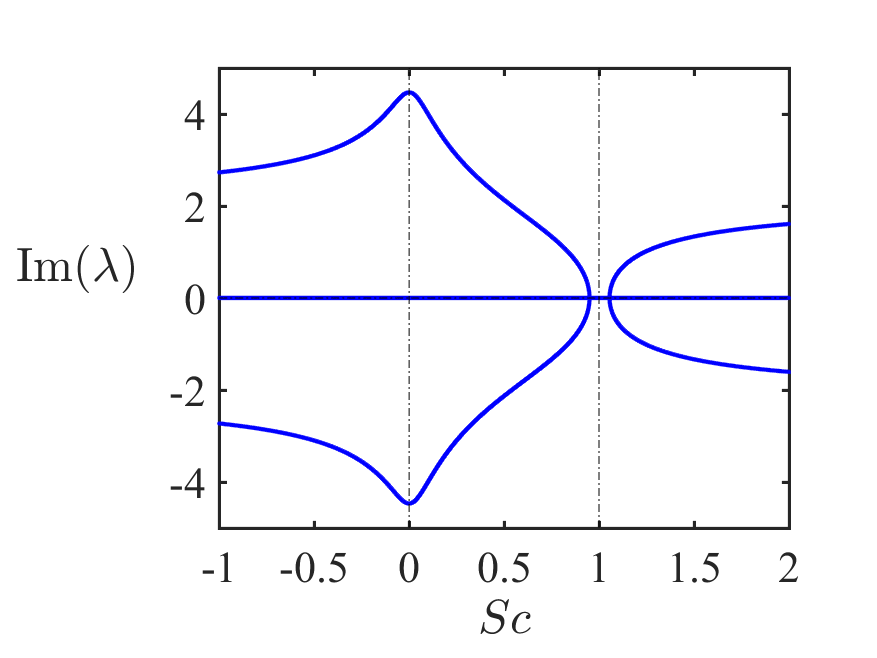}
    \caption{}\label{}
    \end{subfigure}
    
  \caption{Growth rates and frequencies for $m=0$, $Ek=1$, $Ro=1$, and (a,d) $\gamma_1=-4$, $\gamma_2=3$ (b,e) $\gamma_1=-1$, $\gamma_2=3$, (c,f) $\gamma_1=20$, $\gamma_2=-22$, demonstrating exchange of {\color{black}{monotonic}} and oscillatory instabilities near $Sc=1$, cf. Fig.~\ref{fig:codim2g1g2}. \label{fig:grf}}
\end{figure*}

\begin{figure*}[t!]
\centering
    \begin{subfigure}{0.32\textwidth}
    \includegraphics*[width=\textwidth]{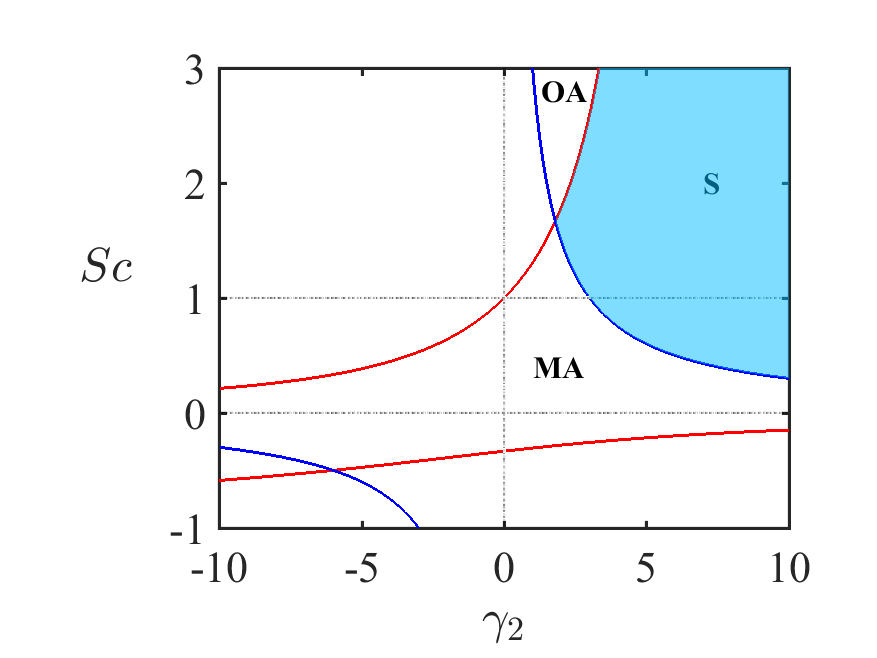}
    \caption{}\label{}
    \end{subfigure}
    \begin{subfigure}{0.32\textwidth}
    \includegraphics*[width=\textwidth]{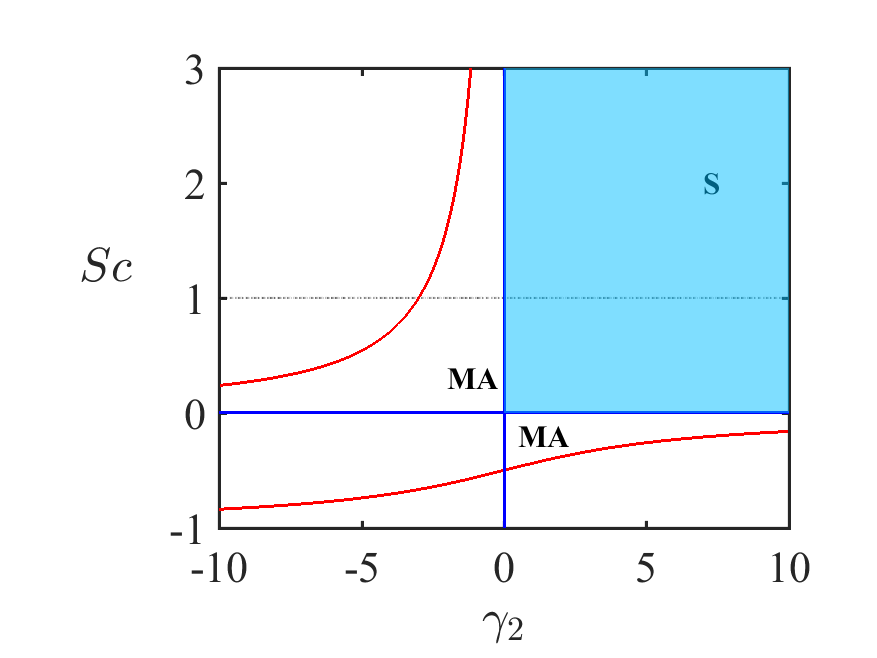}
    \caption{}\label{}
    \end{subfigure}
    \begin{subfigure}{0.32\textwidth}
    \includegraphics*[width=\textwidth]{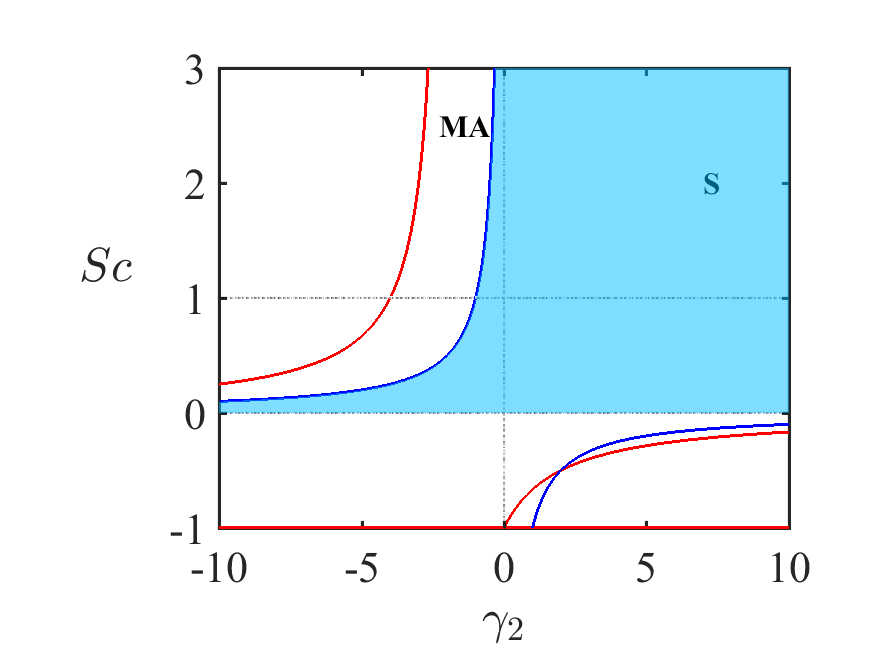}
    \caption{}\label{}
    \end{subfigure}\\
    \begin{subfigure}{0.32\textwidth}
    \includegraphics*[width=\textwidth]{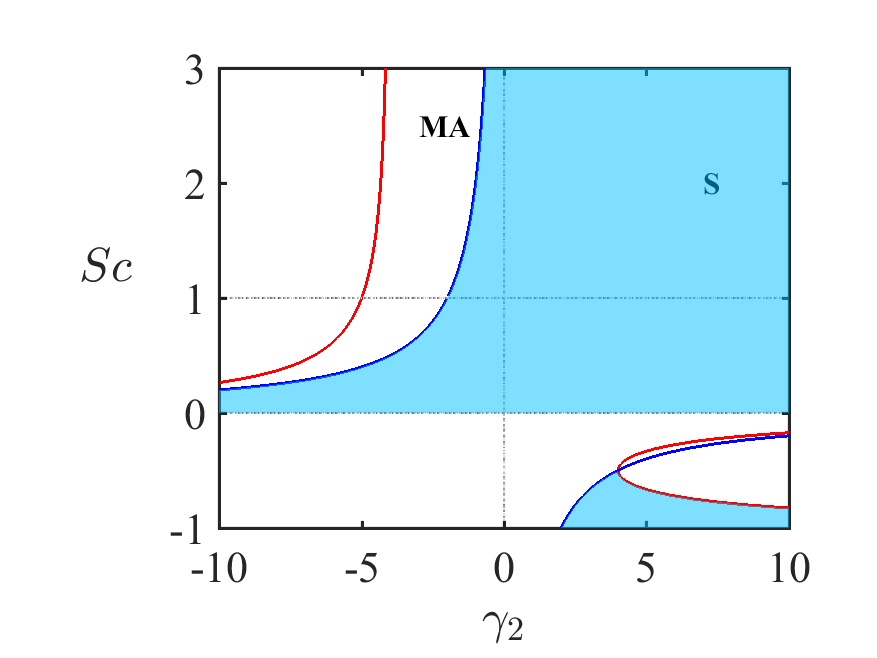}
    \caption{}\label{}
    \end{subfigure}
    \begin{subfigure}{0.32\textwidth}
    \includegraphics*[width=\textwidth]{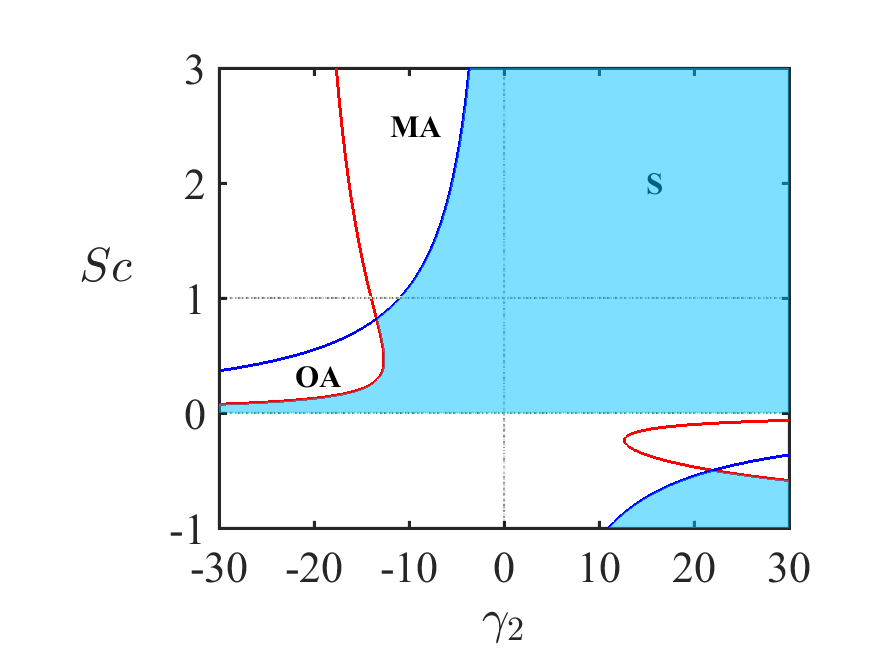}
    \caption{}\label{}
    \end{subfigure}
    \begin{subfigure}{0.32\textwidth}
    \includegraphics*[width=\textwidth]{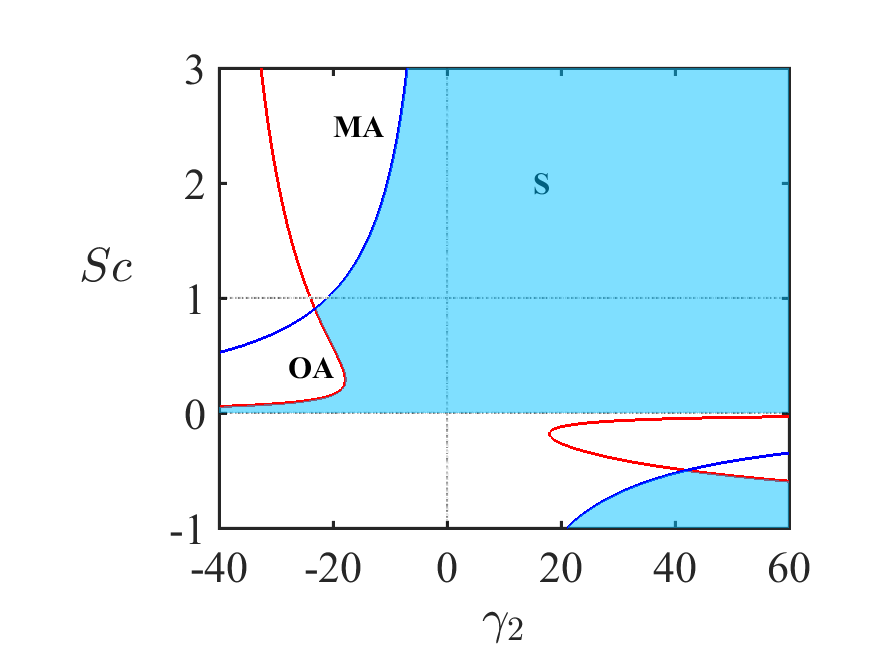}
    \caption{}\label{}
    \end{subfigure}
    
  \caption{Stability maps for $Ek=1$ and (a) $\gamma_1=-4$, (b) $\gamma_1=-1$, (c) $\gamma_1=0$, (d) $\gamma_1=1$, (e) $\gamma_1=10$, (f) $\gamma_1=20$ with the codimension-2 points at (a) $\gamma_2=9/5$ and $Sc=5/3$, (e) $\gamma_2=-121/9$ and $Sc=9/11$, and (f) $\gamma_2=-441/19$ and $Sc=19/21$. At the codimension-2 point $Sc\rightarrow 1$ as $|\gamma_1| \rightarrow \infty$, in accordance with Fig.~\ref{fig:loci}(b). \label{fig:frgrm0}}
\end{figure*}

For $0< Sc < 1$ we have $0>\sigma_1>\sigma_2=-1>\sigma_3$, meaning that the slope of the line \rf{axisym3} is steeper than the slope of \rf{axisym1}, see Fig.~\ref{fig:codim2g1g2}(a-c). Therefore, the neutral stability lines forming the boundary of the stability domain intersect each other at the point \rf{codim2} such that $\gamma_1>0$ and $\gamma_2<0$, see Fig.~\ref{fig:loci}(a). This singular point on the stability boundary is widely known in the hydrodynamical literature as a \textit{codimension-2 point} \cite{KM2017} or Bogdanov-Takens bifurcation point \cite{T2001}. Stability domain is therefore convex, with its edge lying in the domain of centrifugal instability of the diffusionless vortex, Fig.~\ref{fig:codim2g1g2}(a-c). On the other hand, difference of the slopes $\sigma_1$ and $\sigma_3$ from $\sigma_2=-1$ allows for diffusive destabilization of centrifugaly-stable vortices, if the absolute values of $\gamma_1$ and $\gamma_2$ are large enough, Fig.~\ref{fig:codim2g1g2}(a-c).

As $Sc$ approaches 1, the difference between slopes \rf{slopes} is decreased so that $\sigma_1=\sigma_2=\sigma_3=-1$ at $Sc=1$, Fig~\ref{fig:codim2g1g2}(d). This process is accompanied by the movement of the codimension-2 point on the lower branch of the curve \rf{pg1g2} from $\gamma_2 \rightarrow -\infty$ along the asymptotic direction $\gamma_1= Ek^2$  as $Sc$ departs from zero to $\gamma_1 \rightarrow +\infty$ and $\gamma_2\rightarrow -\infty$ along the asymptotic direction $\gamma_1+\gamma_2+3 Ek^2=0$ as $Sc \rightarrow 1$, Fig.~\ref{fig:loci}. 

At $Sc=1$ the stability boundaries of the diffusionless system and the double-diffusive system exactly coincide in the limit of $Ek \rightarrow 0$. However, for $Ek\ne 0$, double diffusion can stabilize centrifugally unstable diffusionless vortices, quite in agreement with Lazar et al. \cite{L2013}, Fig.~\ref{fig:codim2g1g2}(d).

As soon as the Schmidt number passes the threshold $Sc=1$, the codimension-2 point re-appears at infinity on the upper branch of the curve \rf{pg1g2} and moves along the asymptotic direction $\gamma_1+\gamma_2+3 Ek^2=0$ until it reaches a minimum of this curve at $\gamma_1=-Ek^2$ and $\gamma_2=0$ when $Sc \rightarrow +\infty$, Fig.~\ref{fig:loci}. This qualitative change in location of the codimension-2 point (cf. Tuckerman \cite{T2001}) is accompanied by the exchange of the stability criteria: the condition \rf{axisym3} becomes dominating over \rf{axisym1} and vice versa, Fig.~\ref{fig:codim2g1g2}(e,f).  Notice that $Sc=700$ in recent experiments \cite{Z2017}.

\begin{figure*}[t!]
\centering
    \begin{subfigure}{0.4\textwidth}
    \includegraphics*[width=\textwidth]{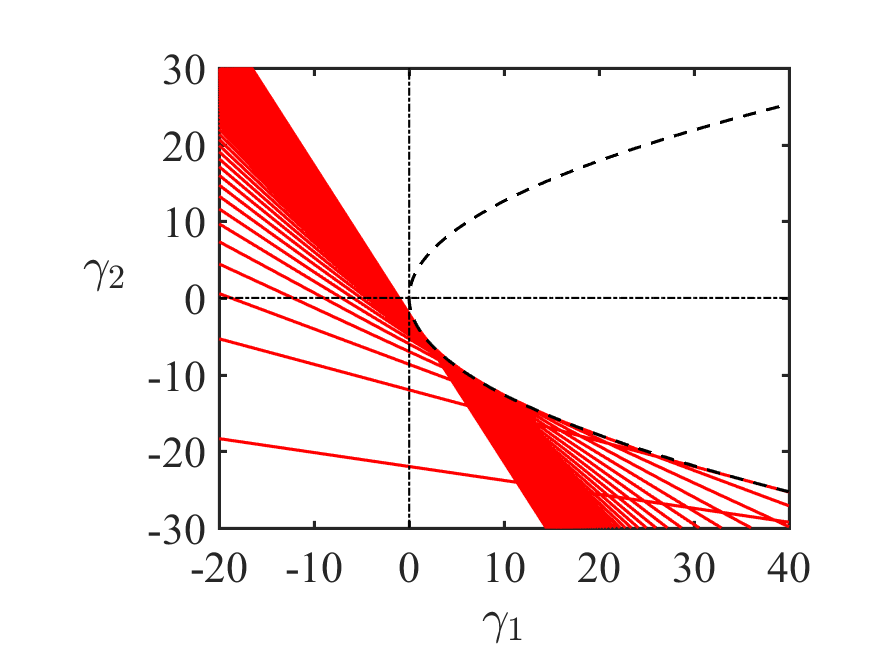}
    \caption{}\label{}
    \end{subfigure}
    \begin{subfigure}{0.4\textwidth}
    \includegraphics*[width=\textwidth]{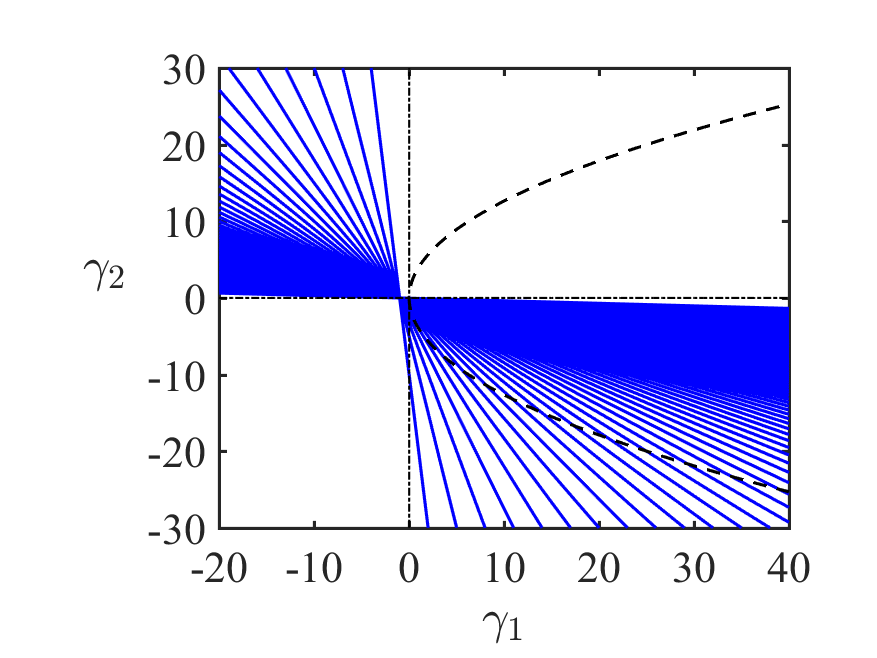}
    \caption{}\label{}
    \end{subfigure}
    
  \caption{(Dashed) For $Ek=1$ the parabolic envelope \rf{envelop}  of {\color{black}{(a)}} a family of straight lines \rf{axisym1}, parameterized by $0<Sc<30$, that determine the boundary between the domain of stability and oscillatory axisymmetric instability. Inside the parabolic region there is no oscillatory axisymmetric instability for all $Sc>0$. {\color{black}{(b)}} A family of straight lines \rf{axisym3} parameterized by $0<Sc<30$, that determine the boundary between the domain of stability and {\color{black}{monotonic}} axisymmetric instability; all the lines in this family have a common point at $\gamma_1=-Ek^2$ and $\gamma_2=0$. \label{fig:caustic}}
\end{figure*}

\subsection{Exchange of {\color{black}{monotonic}} and oscillatory instabilities}

Actually, reversed inequality \rf{axisym3} determines {\color{black}{monotonic}} axisymmetric {\color{black}{(MA)}} instability, corresponding to a monotonically growing perturbation, while the reversed inequality \rf{axisym1} stands for oscillatory axisymmetric (OA) instability, i.e. growing oscillation.

Fig.~\ref{fig:codim2g1g2} provides evidence that stability boundary consisting of two straight lines that intersect at a codimension-2 point in the $(\gamma_1,\gamma_2)$-plane exhibit a qualitative change at $Sc=1$ such that for $Sc<1$ ($Sc>1$) the upper (lower) line corresponds to the onset of SA and the lower (upper) line to the onset of OA. Notice that, as described above, the location of the codimension-2 point changes with the change of $Sc$  with a `jump' at $Sc=1$.

By the latter reason, the qualitative fact of exchange of {\color{black}{monotonic}} and oscillatory axisymmetric instabilities at $Sc=1$, so evident in the $(\gamma_1,\gamma_2)$-plane, is obscured in the plots of growth rates and frequencies of the perturbation versus $Sc$.

Indeed, in Fig.~\ref{fig:grf}(a,d) $\gamma_1=-4$, exactly as in Fig.~\ref{fig:frgrm0}(a), where a codimension-2 point exists at $\gamma_2=9/5$ and $Sc=5/3>1$, separating the boundaries of {\color{black}{monotonic}} ($Sc<5/3$) and oscillatory ($Sc>5/3$) axisymmetric instabilities. Although the growth rates and frequencies in Fig.~\ref{fig:grf}(a,d)  computed at $\gamma_2=3$ confirm the order of SA and OA, the exchange between these instabilities occurs not exactly at $Sc=1$, but in a neighborhood of this value. According to Fig.~\ref{fig:loci}(b) the critical value of $Sc \rightarrow 1$ as $\gamma_1 \rightarrow -\infty$.

Changing the sign of $\gamma_1$ from negative to positive leads to re-appearance of the codimension-2 point  in the second quadrant in the $(\gamma_2, Sc)$-plane, Fig.~\ref{fig:frgrm0}(e,f). In particular, for $\gamma_1=20$ it is situated at $\gamma_2=-441/19$ and $Sc=19/21<1$. The codimension-2 point separates the boundaries of oscillatory ($Sc<19/21$) and {\color{black}{monotonic}} ($Sc>19/21$) axisymmetric instabilities that are in the reverse order with respect to the case of negative $\gamma_1$. Again, growth rates and frequencies computed in Fig.~\ref{fig:grf}(c,f) for $\gamma_2=-22$, confirm that transition from OA to SA occurs at a value of $Sc$ in the vicinity of $Sc=1$. The critical value of $Sc$ tends to 1 as $\gamma_1 \rightarrow +\infty$ in agreement with Fig.~\ref{fig:loci}(b).

We notice that according to \rf{axisym1}--\rf{axisym3}, the described qualitative picture with destabilization of centrifugally stable vortices for $Sc\ne 1$, codimension-2 point, and exchange of instabilities is preserved even in the limit of vanishing dissipation, $Ek \rightarrow 0$, in accordance with the properties of the McIntyre instability \cite{M1970}, which are typical for a broad class of dissipation-induced instabilities \cite{KV2010,K2021}.

\subsection{OA as a genuine dissipation-induced instability}

In Fig.~\ref{fig:codim2g1g2} and Fig.~\ref{fig:frgrm0} one can see that the codimension-2 point separates the boundaries of the regions of oscillatory and {\color{black}{monotonic}} axisymmetric instabilities. The existence of the codimension-2 point qualitatively distinguishes the diffusive case from the diffusionless one, where the onset of instability corresponds to the {\color{black}{monotonic}} axisymmetric centrifugal instability only. 

The growth rate of the oscillatory instability is smaller than the growth rate of the centrifugal instability, Fig.~\ref{fig:grf}. However, in contrast to McIntyre \cite{M1970}, who found such modes within the domain of centrifugal instability and concluded that they are not important with respect to centrifugally-unstable modes that are always destabilized first in his setting, we discovered the conditions when the oscillatory axisymmetric modes are destabilized first and thus determine the onset of instability. 

Hence, the oscillatory axisymmetric instability is a genuine dissipation-induced instability \cite{KV2010,K2021} which is as important as the {\color{black}{monotonic}} axisymmetric one despite its relatively low growth rate, because in a large set of parameters the oscillatory axissymmetric modes are the first to be destabilized by the differential diffusion of mass and momentum.

\subsection{Sufficient conditions for the vortex stability at any $Sc>0$}

Notice that the family of straight lines given by equating to zero the left-hand side of the inequality \rf{axisym1} and parameterized with $Sc$ has a non-trivial envelope, see Fig.~\ref{fig:caustic}(a). To find it, we differentiate the left-hand side by $Sc$, express $Sc$ from the result to substitute it back to \rf{axisym1}. This yields the following parabola in the $(\gamma_1,\gamma_2)$-plane
\be{envelop}
\gamma_1 = \frac{\gamma_2^2}{16Ek^2},
\ee
shown as a dashed curve in Fig.~\ref{fig:codim2g1g2}, Fig.~\ref{fig:loci}, and Fig.~\ref{fig:caustic}. One can see that as $Sc$ increases from 0 to infinity, the OA-boundaries are accumulating and ultimately tend to the line
\be{oaline}
\gamma_2=-2 (Ek^2 +\gamma_1)
\ee
that passes through the point $\gamma_1=-Ek^2$ and $\gamma_2=0$ in the $(\gamma_1,\gamma_2)$-plane, see Fig.~\ref{fig:caustic}(a). 

This implies that in all the points inside the parabolic envelope \rf{envelop} the vortex cannot be destabilized via the oscillatory instability mechanism, no matter what is the value of $Sc>0$. To the best of our knowledge this explicit result has never been reported in the literature. 

On the other hand, the family of straight lines given by equating to zero the left-hand side of the inequality \rf{axisym3} varies between the line $\gamma_1=-Ek^2$ at $Sc=0$ and the line $\gamma_2=0$ at $Sc\rightarrow \infty$, Fig.~\ref{fig:caustic}(b). Therefore, the whole lower part of the parabola \rf{envelop} can belong to the domain of {\color{black}{monotonic}} axisymmetric instability in the limit of infinite $Sc>0$.

Consequently, the area in the $(\gamma_1,\gamma_2)$-plane, limited by the criteria 
\be{sufco}
\gamma_1\ge -Ek^2, \quad \gamma_2\ge 0
\ee
corresponds to the stability domain, no matter what is the value of $Sc>0$. This is in agreement with the analysis of the case $\gamma_2=0$ based on the equations \rf{g20} and generalises the results of McIntyre \cite{M1970} and Singh and Mathur \cite{SM2019} due to more comprehensive structure of parameters $\gamma_1$ and $\gamma_2$ given by \rf{gamma}.

\subsection{Non-axisymmetric case}

{\color{black}{Finally, we notice that the stability defined by the set of inequalities \rf{bilh1}-\rf{bilh3} obtained from the Bilharz criterion does not exhibit dependence on the azimuthal wavenumber $m$, despite the right sides of the expressions do contain $m$ explicitly, making the non-axisymmetric and axisymmetric criteria match exactly. 

We may emphasize that we limited ourselves by the lowest-order in $\varepsilon$ terms in the asymptotic expansion of expressions \rf{goa} that in this particular class of problems has led to the dispersion relation that does not distinguish between the neutral stability curves of axisymmetric and non-axisymmetric instabilities, although it was capable to catch non-axisymmetric instabilities in the same order approximation, e.g., in the studies of magnetorotational instability \cite{KSF2014,K2017}.

On the other hand, as we have seen from the literature review in the Introduction, axisymmetric instabilities is a prevailing type of instabilities in the studies of circularly symmetric vortices and, perhaps, due to the symmetry the non-axisymmetric instabilites will reveal itself in the next-order terms of the geometric optics method.}}

\section{Conclusion}

We considered a model of a baroclinic circular lenticular vortex with a Gaussian profile of angular velocity both in radial and axial directions, immersed in a vertically stratified viscous fluid in the presence of diffusion of a stratifying agent and rotation of the coordinate frame related to the ambient fluid. This setting is substantially more comprehensive than those of the previous works that, in particular, were limited by the assumption of barotropy, did not take into account rotation of the frame and diffusion of mass and momentum, or set the Schmidt number equal to unity. 

We have derived an original dimensionless set of equations on the $f$-plane, describing the dynamics of the vortex immersed in a vertically stratified fluid and then linearized it about a base state that we have found explicitly. The linearized equations of motion were further expanded in terms of asymptotic series by means of the geometric optics approximation \cite{KSF2014,KM2017,K2017,SM2019,Vidal2019,K2021} to produce a set of the amplitude transport equations. The latter offered us an opportunity to derive an exhaustive but elegant third-order polynomial dispersion relation governing the local stability of the vortex. 

In the diffusionless limit and in the case where magnitudes of both damping mechanisms are identical we obtained a generalized Rayleigh criterion for centrifugal instability in terms of the shear and buoyancy parameters $\gamma_1$ and $\gamma_2$ and shown that it reduces to the known in the literature particular cases.

Applying the algebraic Bilharz criterion to the complex dispersion relation we derived
new rigorous stability criteria in terms of $\gamma_1$ and $\gamma_2$ as well as the Schmidt and Ekman numbers related to the differential diffusion of mass and momentum. We visualized these criteria in the $(\gamma_1,\gamma_2)$-plane and revealed  a codimension-2 point splitting the boundaries of oscillatory and {\color{black}{monotonic}} axisymmetric instabilities that can affect both centrifugally stable and unstable diffusionless flows. 

The oscillatory axisymetric instability was found to be a genuine dissipation-induced instability because of its absence in the diffusionless case. Nevertheless, we have described explicitly a parabolic region in the $(\gamma_1, \gamma_2)$-plane that is free of oscillatory axisymmetric instabilities, no matter what the value of $Sc>0$ is.

In contrast to the work of McIntyre \cite{M1970} we found conditions when oscillatory axissymmetric modes are the first to be destabilized by the double diffusion and thus are dominant even despite the growth rate of the oscillatory instability is generally weaker than that of the centrifugal instability \cite{YB2016}. Finally, we provided a sufficient condition for stability of a baroclinic vortex at arbitrary $Sc>0$ that generalizes that of the previous works by McIntyre \cite{M1970} and Singh and Mathur\cite{SM2019}. 

This study conclusively proved the decisive role of the Schmidt number and therefore the differential diffusion of mass and momentum for the stability of lenticular vortices and, particularly, for excitation of the genuine dissipation-induced oscillatory instability. A codimension-2 point found on the neutral stability curve is proven to govern exchange of {\color{black}{monotonic}} and oscillatory instability as the Schmidt number transits through the unit value. 
All the results are preserved even in the limit of vanishing dissipation, which is a typical property of dissipation-induced instabilities \cite{KV2010,K2021}.

We have thus developed new analytical criteria for an express-analysis of stability of baroclinic circular lenticular vortices for arbitrary parameter values that is believed to be an efficient tool for informing future numerical and experimental studies in this actively developing field.

\section*{Acknowledgements}
J.L. was supported by a Ph.D. Scholarship from Northumbria University and by a Postdoctoral Fellowship from the Institut de M{\'e}canique et d'Ing{\'e}nierie at Aix-Marseille Universit{\'e}. The research of O.N.K. was supported in part by the Royal Society grant $\rm IES \backslash R1 \backslash 211145$.

\nocite{*}
\bibliography{aipsamp}

\end{document}